\documentclass[aps,pra,amsmath,amsfonts,amssymb,twocolumn,showpacs]{revtex4-1}

\usepackage{graphicx,psfrag,color}% Include figure files
\usepackage{dcolumn}% Align table columns on decimal point
\usepackage{bm}% bold math
\usepackage{mathbbol}

\newcommand{\im}{\mathrm i}

\newcommand{\med}[1]{\left\langle #1 \right\rangle}
\begin{document}

\title{Spotlighting quantum critical points via quantum correlations at 
finite temperatures}

\author{T. Werlang}
%\affiliation{Departamento de F\'{i}sica,
%Universidade Federal de S\~ao Carlos, S\~ao Carlos, SP 13565-905,
%Brazil}
\author{G.A.P. Ribeiro}
%\affiliation{Departamento de F\'{i}sica,
%Universidade Federal de S\~ao Carlos, S\~ao Carlos, SP 13565-905,
%Brazil}
\author{Gustavo Rigolin}
\email{rigolin@ufscar.br}
\affiliation{Departamento de F\'{i}sica,
Universidade Federal de S\~ao Carlos, S\~ao Carlos, SP 13565-905,
Brazil}

\date{\today}

\begin{abstract}
We extend the program initiated in [T. Werlang \textit{et al.}, Phys. Rev.
Lett. \textbf{105}, 095702 (2010)] in several directions.
Firstly, we investigate how useful quantum correlations, such as entanglement
and quantum discord, are in the detection 
of critical points of quantum phase transitions when the system is 
at finite temperatures. 
For that purpose we study several thermalized spin models in the 
thermodynamic limit, namely, the XXZ model, the
XY model, and the Ising model, all of which with an external magnetic field.
We compare the ability of quantum discord, entanglement, and some thermodynamic
quantities to spotlight the quantum critical points for several different 
temperatures. Secondly, for some models we go beyond nearest-neighbors and 
also study the behavior of 
entanglement and quantum discord for second nearest-neighbors around the 
critical point at finite temperature. Finally, we furnish a more quantitative
description of how good all these quantities are in spotlighting critical
points of quantum phase transitions at finite $T$, bridging the gap between
experimental data and those theoretical descriptions solely based on the 
unattainable absolute zero assumption.
\end{abstract}


\maketitle

\section{Introduction}

Quantum phase transitions (QPTs) theoretically occur at absolute zero 
temperature $(T=0)$ due to abrupt changes in the qualitative properties of the 
ground-state of a many-body system while varying its Hamiltonian 
\cite{Sac99}. The tuning parameter (the quantity being changed in the 
Hamiltonian) can be, for example, an external magnetic field or a coupling 
constant. By properly tuning the Hamiltonian one can reach a special point, 
the critical point (CP), where the ground state of the system undergoes a 
radical change, which strongly affects the macroscopic properties of 
the system. In the vicinity of a CP, a tiny change in the tuning parameter 
will favor one phase over the other. Since at $T=0$ there are no thermal 
fluctuations,  these tiny changes are caused only by quantum fluctuations, 
whose origin can be traced back to the Heisenberg uncertainty principle. 
Some well-known examples of QPTs are the paramagnetic-ferromagnetic transition 
in some metals \cite{rowley}, the superconductor-insulator transition 
\cite{dolgopolov}, and superfluid-Mott insulator transition \cite{greiner}.

Although reaching the absolute zero is impossible since it would 
violate the third law of thermodynamics, QPTs can be detected if one reaches
temperatures near the absolute zero. More precisely, when the system's 
de Broglie wavelength is greater than the correlation length of the thermal 
fluctuations one achieves a temperature low enough to see a QPT. 
In this regime thermal fluctuations are unable to excite the system from its 
ground state and a phase transition solely driven by quantum fluctuations is 
still possible to be seen.

Many theoretical tools employed in the calculation of these CPs are usually 
subjected to the $T=0$ restriction. Interesting examples are the 
behavior of bipartite \cite{lidar}, multipartite \cite{oliveira}, and
generalized entanglement \cite{ortiz} for spin chains at $T=0$. Quantum 
discord, a measure of all quantum correlations 
present in a system \cite{zurek,Henderson}, has also been widely used to 
characterize the CPs at $T=0$ \cite{Dil08,Sar09}. 
The extremal values and the behavior of the 
derivatives of these quantities can signal the existence of a CP without the
knowledge of an order parameter, i.e., without the knowledge of what 
macroscopic quantity abruptly changes during the QPT.

All experiments involving QPTs are, nevertheless, 
performed at low temperatures, 
not at absolute zero, which forbid a direct connection between the measured 
data and those theoretical results developed for $T=0$. Recently, however,
we have studied, in the thermodynamic limit, the XXZ Hamiltonian with no 
magnetic field
and shown that quantum discord (QD) is able to detect CPs at finite $T$ 
while other thermodynamic quantities and entanglement fail in this 
task \cite{werPRL}. For that model we have shown that the thermal quantum
discord (TQD), i.e., QD computed for a system described by the canonical 
ensemble, possessed a characteristic behavior at the CP robust enough to 
survive an increasing $T$ and, therefore, spotlight a CP of a QPT at 
finite $T$.

Our goal in this work is to extend the study initiated in 
\cite{werPRL} in at least two directions.  We want, on one hand,
to study different models to see whether or not TQD is still a good CP 
detector at finite $T$. We want to answer the following question: Does
TQD possess a clear distinctive behavior at the CP strong enough to be
seen at $T>0$? On the other hand, we want to be more quantitative in 
determining the ability of TQD, and other quantities such as
entanglement, to correctly detect the CP of a QPT at finite $T$. 
How close to the actual
CP are the estimated CPs obtained at finite $T$ via TQD? How do these 
estimated CPs at $T>0$ approach the correct CP as we decrease $T$? Those
are the questions we also want to answer in this paper.

To achieve these goals, we consider here the XXZ model in the presence of an 
external magnetic field and the XY model in a transverse field. For both 
models we analyze the behavior of TQD and entanglement for 
$T\geq0$ around their CPs. For the XXZ model with an external field we 
compare the efficiency of TQD with respect to entanglement in the detection 
of CPs as the temperature increases and also with 
other thermodynamic quantities. For the XY model we study the behavior of
 TQD and the entanglement of its first and second nearest-neighbors as a CP 
detector.

In order to make this work self-contained, we organize its presentation as
follows. In Sec. II we review some concepts from 
information theory necessary to the construction of two measures of quantum 
correlations: QD and entanglement of formation (EoF). 
In Sec. III we discuss qualitatively and quantitatively 
the role of quantum correlations as a tool to 
detect QPTs at zero and finite temperatures in the XXZ and XY models, both
with external magnetic fields. Our final remarks and conclusions are 
presented in Sec IV. 

\section{Quantum Correlations}

\subsection{Quantum Discord}

In the context of classical information theory \cite{nielsen}, 
correlation is a measure of the dependence between two or more random 
variables or observed data values. The total correlation between two 
random variables $A$ and $B$ with probability distribution $p_a$ and $p_b$, 
respectively, is given by the mutual information (MI) \cite{nielsen}
\begin{eqnarray}\label{mi1}
\mathcal{I}_1(A:B)= \mathcal{H}(A)+\mathcal{H}(B)-\mathcal{H}(A,B),
\end{eqnarray}
where $\mathcal{H}(X)=-\sum_kp_x\log_2p_x$ is the Shannon entropy with $p_x$ 
the probability distribution of the random variable $X$. If $p_{a,b}$ denote 
the joint probability distribution of the variables $A$ and $B$ then the joint 
entropy $\mathcal{H}(A,B)$ can be written through the Bayes' rule 
$p_{a|b}=p_{a,b}/p_{b}$ \cite{bayes} as 
$\mathcal{H}(A,B)=\mathcal{H}(A|B)+\mathcal{H}(B)$, where the conditional 
entropy $\mathcal{H}(A|B)=-\sum_{a,b}p_{a,b}\log_2p_{a|b}$ quantifies how much 
uncertainty is left on average about $A$ when one knows the values of $B$. 
Note that the conditional entropy is always non-negative and not necessarily a 
symmetric quantity, i.e., $\mathcal{H}(A|B)$ may be different from 
$\mathcal{H}(B|A)$. Using this relation between $\mathcal{H}(A,B)$ and 
$\mathcal{H}(A|B)$ we can obtain an equivalent version for the MI as
given in (\ref{mi1}), 
\begin{eqnarray}\label{mi2}
\mathcal{I}_2(A:B)= \mathcal{H}(A)-\mathcal{H}(A|B).
\end{eqnarray}

For a quantum bipartite system $AB$ described by the density operator 
$\rho_{AB}$, a measure of their total correlation is obtained directly from 
Eq.~(\ref{mi1}) replacing the Shannon entropy by the von-Neumann entropy, 
$\mathcal{S}(X)=\mathcal{S}(\rho_X)=-\mbox{Tr}\left(\rho_X\log_2\rho_X\right)$,
i.e., $\mathcal{I}^q_1(A:B)= \mathcal{S}(A)+\mathcal{S}(B)-\mathcal{S}(A,B)$. 
However, a quantum analog of the second version of MI, Eq.~(\ref{mi2}), 
cannot be obtained so easily because Bayes' rule are not always valid in the 
quantum case \cite{peres}. 
For example, assuming the validity of this rule, the quantum conditional 
entropy would be defined as 
$\mathcal{S}(A|B)=\mathcal{S}(A,B)-\mathcal{S}(B)$. But this 
definition implies a negative value for $\mathcal{S}(A|B)$ if one
computes it for the quantum state 
$\left|\psi\right\rangle=\left(\left|00\right\rangle+
\left|11\right\rangle\right)/\sqrt{2}$, showing that this quantity 
cannot be interpreted as in the classical case, where it is always positive.   

To ensure that the quantum conditional entropy has the same meaning as in 
the classical case, Henderson and Vedral \cite{Henderson} defined the 
conditional entropy as 
$\mathcal{S}_q(A|B)\equiv\min_{\left\{M_b\right\}}\sum_bp_b\mathcal{S}(A|B=b)$ 
such that the minimization is given over generalized 
measurements $\left\{M_b\right\}$ \cite{nielsen}, 
with $\sum_bM_b=\boldsymbol{1}_B$, $\boldsymbol{1}_B$ the identity operator
that acts on $B$, $M_b\geq0$ for all $b$, and 
$\mathcal{S}(A|B=b)=\mathcal{S}(\rho_{A|b})$ where
\begin{eqnarray*}
\rho_{A|b}=\frac{1}{p_b}(\boldsymbol{1}_A\otimes M_b)\rho_{AB}(\boldsymbol{1}_A
\otimes M_b),
\end{eqnarray*} 
with $p_b=\mbox{Tr}\left[(\boldsymbol{1}_A\otimes M_b)
\rho_{AB}(\boldsymbol{1}_A\otimes M_b)\right]$. 
The expressions $\mathcal{S}_q(A|B)$ and $\mathcal{S}(A|B)$ coincide with 
$\mathcal{H}(B|A)$ for classical systems. Using $\mathcal{S}_q(A|B)$ the 
second quantum version for MI, Eq.~(\ref{mi2}), can be written as 
$\mathcal{I}^q_2(A:B)= \mathcal{S}(A)-\mathcal{S}_q(A|B)$. The difference 
between these two versions of quantum mutual information,   
\begin{eqnarray}\label{qd}
D(A|B)&\equiv&\mathcal{I}^q_1(A:B)-\mathcal{I}^q_2(A:B),\nonumber\\
&=&\mathcal{S}(B)-\mathcal{S}(A,B)+\mathcal{S}_q(A|B),
\end{eqnarray}
was called 
{\it quantum discord} by Ollivier and Zurek \cite{zurek}
and interpreted as a measure of total quantum correlation. 
The notation $D(A|B)$ indicates that the measurements were performed on 
the subsystem $B$. When $\mathcal{S}(A)=\mathcal{S}(B)$ we have 
$D(A|B)=D(B|A)=D\left(\rho_{AB}\right)$ and therefore the QD is a symmetric 
quantity. As demonstrated recently \cite{ved01} the state $\rho_{AB}$ has 
$D(A|B)=0$ if, and only if, there exists a von Neumann measurement 
$\left\{\Pi_j^B\right\}$ such that 
$\rho_{AB}=\sum_j \left(\boldsymbol{1}_A\otimes\Pi_j^B\right)\rho_{AB}
\left(\boldsymbol{1}_A\otimes\Pi_j^B\right)$ 
\cite{footnote1}. Therefore, a state $\rho_{AB}$ 
with $D=0$ has necessarily the form 
$\rho_{AB}=\sum_{j,k}p_{j,k} \left|\psi_j\right\rangle\left\langle \psi_j\right| 
\otimes\left|\phi_k\right\rangle\left\langle \phi_k\right|$ with 
$\sum_{j,k}p_{j,k}=1$ and $\left\{\left|\psi_j\right\rangle\right\}$ and 
$\left\{\left|\phi_k\right\rangle\right\}$ sets of orthogonal states. 
The states with $D=0$ are completely classically
correlated in the sense of \cite{cstates}. 
An example of such state is 
$\rho_{AB}=\left(  \left|0\right\rangle\left\langle 0\right|
\otimes\left|0\right\rangle\left\langle 0\right| +
\left|1\right\rangle\left\langle 1\right|\otimes\left|1\right\rangle\left
\langle 1\right| \right)/2$. This result shows that the existence of quantum 
correlations is due to the principle of superposition that allows describing a 
system through a set of non-orthogonal states. Here, and analogously 
to entanglement computed for thermal states \cite{arnesen}, we use the
nomenclature coined in \cite{Wer10}, namely, thermal quantum discord (TQD),
to refer to QD computed for states described by the canonical ensemble. 

The main difficulty in the determination of a closed expression for QD lies 
in the complicated minimization procedure for calculating the conditional 
entropy $\mathcal{S}_q(A|B)$. For two-qubit systems the minimization over 
generalized measurements is equivalent to a minimization over projective 
measurements (von Neumann measurements) \cite{minDIS}. In this case, the 
minimization procedure can be done numerically for general two-qubit states 
\cite{numDIS} while an analytical expression can be obtained only for a 
subclass of the so-called X-states (see Appendix). In this paper we employed 
both strategies, i.e., numerical minimization and, when available, 
closed expressions for computing QD.

\subsection{Entanglement}

Next, we introduce an important kind of quantum correlated states, 
the entangled states. 
A quantum bipartite state described by $\rho_{AB}$ is said to be 
{\it entangled} if, and only if, it cannot be written as a separable state 
$\rho_{AB}=\sum_jp_j\rho_j^A\otimes\rho_j^B$, where $p_j>0$ and 
$\sum_jp_j=1$ \cite{Wer89,entangle}. 
Although there are states with non-zero quantum discord 
created only by local operations and classical communication (LOCC), 
entanglement cannot be generated in this way \cite{Wer89}. 
For pure states all quantum correlated states are entangled. The situation is 
more complicated in the case of mixed states. For example, the Werner 
state $\rho_w=\alpha\left|\psi^-\right\rangle\left\langle 
\psi^-\right|+(1-\alpha)\boldsymbol{1}/4$, with 
$\left|\psi^-\right\rangle=
\left(\left|01\right\rangle-\left|10\right\rangle\right)/\sqrt{2}$ and 
$\alpha\in[0,1]$, is separable for $\alpha<1/3$ and non-local (violates
a Bell-like inequality) for $\alpha>1/\sqrt{2}$. 
However, the Werner state have null quantum discord only for $\alpha=0$. 

In this paper the measure of entanglement used is the {\it Entanglement of 
Formation} (EoF) \cite{Woo98}, which is defined as 
$EoF(\rho_{AB})=\min\sum_jp_j\mathcal{S}(\rho_{A(B)})$, 
where the minimization is over all ensembles of pure states 
$\left\{p_j,\left|\psi_j\right\rangle\right\}$ such that 
$\rho_{AB}=\sum_jp_j\left|\psi_j\right\rangle\left\langle \psi_j\right|$ and 
$\mathcal{S}(\rho_{A(B)})$ is the von Neumann entropy of 
the reduced state of either $A$ or $B$, 
$\rho_{A(B)}=\mbox{Tr}_{B(A)}\left(\rho_{AB}\right)$. 
The EoF quantifies, at least for pure states
and asymptotically, how many singlets are needed per copy to prepare many 
copies of $\rho_{AB}$ using only LOCC. It is worth emphasizing that it
 coincides with the quantum discord when one deals with pure states. 
An analytic closed expression for EoF is given in the Appendix. 

\section{Results and Discussions}
\label{results}

\subsection{The XXZ Model}
\label{secXXZ}

The first model tackled in this paper is the one-dimensional anisotropic 
spin-$1/2$ Heisenberg chain (XXZ) subjected to an external magnetic field in 
the $z$-direction. The Hamiltonian of this model can be written as 
\begin{eqnarray}\label{hxxz}
H_{xxz}&=&J\sum_{j=1}^L\left(\sigma_j^x\sigma_{j+1}^x+\sigma_j^y\sigma_{j+1}^y+
\Delta\sigma_j^z\sigma_{j+1}^z\right)\nonumber\\ &-&\frac{h}{2}
\sum_{j=1}^L\sigma_j^z,
\end{eqnarray}
where $\sigma_j^\alpha$ ($\alpha=x,y,z$) are the usual Pauli matrices acting on 
the $j$-th site and $\sigma_{L+1}^\alpha=\sigma_1^\alpha$ (periodic boundary 
conditions). Here $h$ is the external magnetic field, $J$ is the 
exchange constant ($J=1$) and 
$\Delta$ is the anisotropy parameter. We are going to consider only the 
case $\Delta>0$, however the negative values $\Delta<0$ can be obtained by 
reversing the sign of  $J$ followed by a canonical transformation \cite{YANG}. 
The critical regime $|\Delta| \leq 1$ can be parametrized by 
$\Delta=\cos{\gamma}$  and the non-critical regime $|\Delta| > 1$ is given 
by $\Delta=\cosh{\gamma}$.
The thermodynamical properties are obtained  through the free-energy 
$f=-\frac{1}{\beta} \lim_{L\rightarrow \infty} \frac{\ln{Z}}{L} $, 
where $Z=\mbox{Tr}\left\{\exp{\left(-\beta H_{xxz}\right)}\right\}$ is the 
partition function, with $\beta=1/kT$ and the Boltzmann constant $k$ is set 
to unity.

The free-energy is calculated by Bethe ansatz techniques and can be written as 
follow,
\begin{equation}
f=e_0 - \frac{1}{\beta} \left( V \ast \ln{B\bar{B}} \right)(0),
\end{equation}
where the ground state energy $e_0$ is given by
\begin{equation}
\frac{e_0}{J}=
\begin{cases}
\cos{\gamma}- 2  \frac{\sin{\gamma}}{\gamma} \int_{-\infty}^{\infty}
\frac{\sinh{(\frac{\pi}{\gamma}-1)\frac{k}{2}}}{2\sinh{\frac{\pi k}{2\gamma}}
\cosh{\frac{k}{2}}}dk, & 0< \Delta \leq 1, \\
\cosh{\gamma} -2  \sinh{\gamma}\sum_{k=-\infty}^{\infty} 
\frac{e^{-\gamma |k|}}{\cosh{\gamma k}}, & \Delta > 1,
\end{cases}
\end{equation}
and
\begin{equation}
 V(x)=\begin{cases}
 \frac{\pi}{\cosh{ \pi x}}, & 0< \Delta \leq 1,  \\
 \sum_{k=-\infty}^{\infty}\frac{e^{\im 2 k x}}{2\cosh{\gamma k}}, & \Delta>1.
\end{cases}
\end{equation}
The symbol $\ast$ denotes convolution $f*g(x)=\int_{-a}^{a} f(x-y)g(y)dy$, 
where $a\rightarrow \infty$ for $0< \Delta \leq 1$ and $a=\pi/2$ 
for $\Delta>1$.  

The auxiliary functions $b(x)$, $\bar{b}(x)$, and its simply related functions
$B(x)=b(x)+1$ and $\bar{B}(x)=\bar{b}(x)+1$ are solution to the following set 
of non-linear integral equations \cite{KLUMPER92},
\begin{eqnarray}
\ln{b(x)}&=&  d_{+}(x)  +  \left( \!K\!\ast\! \ln{B}\right)\!(x) - 
\left(\! K\!\ast\! \ln{\bar{B}}\right)\!(x+\im\gamma), \\
\ln{\bar{b}(x)}&=& d_{-}(x) + \left(\! K\!\ast\! \ln{\bar{B}}\right)\!(x) - 
\left(\! K \!\ast\! \ln{B}\right)\!(x-\im\gamma).	
\label{NLIE}
\end{eqnarray}
The driving term $d_{\pm}(x)$ is given by
\begin{equation}
d_{\pm}(x)=
\begin{cases}
-2J \beta \frac{\sin{\gamma}}{\gamma} \frac{\pi}{\cosh{ (\pi x/\gamma) }} 
\pm  \frac{\beta h}{2}\frac{\pi}{\pi-\gamma}, & 0< \Delta \leq 1,  \\
-2J \beta \sinh{\gamma} \sum_{k=-\infty}^{\infty}
\frac{e^{\im 2 k x}}{\cosh{\gamma k}} \pm  \frac{\beta h}{2}, & \Delta>1,
\end{cases}
\end{equation}
and the Kernel function
\begin{equation}
K(x)=
\begin{cases}
\int_{-\infty}^{\infty}\frac{\sinh{(\pi-2\gamma)
\frac{k}{2}} e^{\im k x}}{2\sinh{(\pi-\gamma)
\frac{k\gamma}{2}}\cosh{\frac{k}{2}}}dk, & 0< \Delta \leq 1, \\
\sum_{k=-\infty}^{\infty} \frac{e^{-\gamma |k|}}{\cosh{\gamma k}} e^{\im 2 k x}, 
& \Delta > 1.
\end{cases}
\end{equation}

The system density operator $\rho$ is  described 
by the canonical ensemble, i.e., $\rho=\exp{\left(-\beta H_{xxz}\right)}/Z$. 
The nearest-neighbor 
two-spin state is obtained by tracing all but the first two spins, 
$\rho_{1,2}=\mbox{Tr}_{L-2}(\rho)$. The Hamiltonian (\ref{hxxz}) exhibits both 
translational invariance and $U(1)$ invariance 
$\left(\left[H_{xxz},\sum_{j=1}^L\sigma_j^z\right]=0\right)$, therefore the 
reduced density matrix will be given by
\begin{eqnarray}\label{rho}
\rho_{1,2}  = \frac{1}{4}\left(
\begin{array}{cccc}
\rho_{11} & 0 & 0 & 0\\
0 & \rho_{22}  & \rho_{23} & 0 \\
0 & \rho_{23} & \rho_{22} & 0 \\
0 & 0 & 0 & \rho_{44}  \\
\end{array}
\right),
\end{eqnarray}
where
\begin{eqnarray*}
\rho_{11} &=& 1+2\left\langle \sigma^z\right\rangle+\med{\sigma_1^z\sigma_2^z},\\
\rho_{22} &=& 1-\med{\sigma_1^z\sigma_2^z},\\
\rho_{44} &=& 1-2\left\langle \sigma^z\right\rangle+\med{\sigma_1^z\sigma_2^z},\\
\rho_{23} &=& 2\med{\sigma_1^x\sigma_2^x}.
\end{eqnarray*}
The magnetization and the two-point correlations above are obtained in terms 
of the derivatives of the free-energy \cite{NLIE} 
\begin{eqnarray}
\med{\sigma^z}&=& -2\partial_h f, \\
\med{\sigma_j^{z}\sigma_{j+1}^{z}}&=&\partial_{\Delta}f/J,  \\
\med{\sigma_j^{x}\sigma_{j+1}^{x}}&=&\frac{u-\Delta \partial_{\Delta}f+ 
h \med{\sigma^z}}{2J}, \\ 
\med{\sigma_j^{z}\sigma_{j+1}^{z}}&=&\med{\sigma_j^{x}\sigma_{j+1}^{x}}=
\frac{u + h \med{\sigma^z}}{3J}, ~~ \Delta=1,
\end{eqnarray}
where $u=\partial_{\beta} (\beta f)$ is the internal energy.

Using (\ref{rho}) we can obtain analytical expressions for  TQD and EoF 
between two nearest-neighbor spins through Eqs. (\ref{qdX}) and (\ref{eof}), 
respectively. See Appendix for details.
 
As discussed in \cite{TAKAHASHI}, the XXZ model has two CPs in the absence of 
external field ($h=0$). An infinite-order transition at $\Delta=1$, with the 
ground state changing from an XY-like phase ($-1<\Delta<1$) to an Ising-like 
antiferromagnetic phase for $\Delta>1$, and a first-order transition at 
$\Delta=-1$, from a ferromagnetic phase ($\Delta<-1$) to the critical 
antiferromagnetic phase ($-1<\Delta<1$).
In \cite{Dil08,Sar09} 
it was shown that both quantum discord and entanglement are 
able to detect these CPs associated to QPTs at $T=0$. In order to extend these 
zero temperature studies of the XXZ model, we take into account the action 
of the external field $h$. The effect of the external field is to shift the 
critical points to higher values of $\Delta$ \cite{GAUDIN}. The critical 
point associated to the infinite-order transition $\Delta_{inf}$ 
is determined by 
the following equation
\begin{eqnarray}
h=4J\sinh(\eta)\sum_{n=-\infty}^\infty\frac{(-1)^n}{\cosh(n\eta)},
\end{eqnarray}
with $\eta = \cosh^{-1}(\Delta_{inf})$, while the critical point associated to 
the first-order transition $\Delta_1$ by the equation 
\begin{eqnarray}
h=4J(1+\Delta_1).
\end{eqnarray}
Table I shows the critical points associated to first- and infinite-order QPTs 
for different values of the external field $h$.  
\begin{table}[!ht]
\caption{\label{cps}
Critical points associated to the first-order QPT ($\Delta_1$) and to 
the infinite-order QPT ($\Delta_{inf}$) for different values of the external 
field $h$. %The values of $h$ used here were $0$, $2$, $6$, and $12$.
}
\begin{center}
\begin{ruledtabular}
\begin{tabular}{lrrrr}
& $h=0$ & $h=2$ & $h=6$ & $h=12$\\ \hline 
$\Delta_1$ (first-order) & -1.00 & -0.50 & 0.50 & 2.00\\ %\hline
$\Delta_{inf}$ (infinite-order) & 1.00 & 2.15 & 3.30 & 4.88\\
\end{tabular}
\end{ruledtabular}
\end{center}
\end{table}

\begin{figure}[!ht]
\includegraphics[width=8cm]{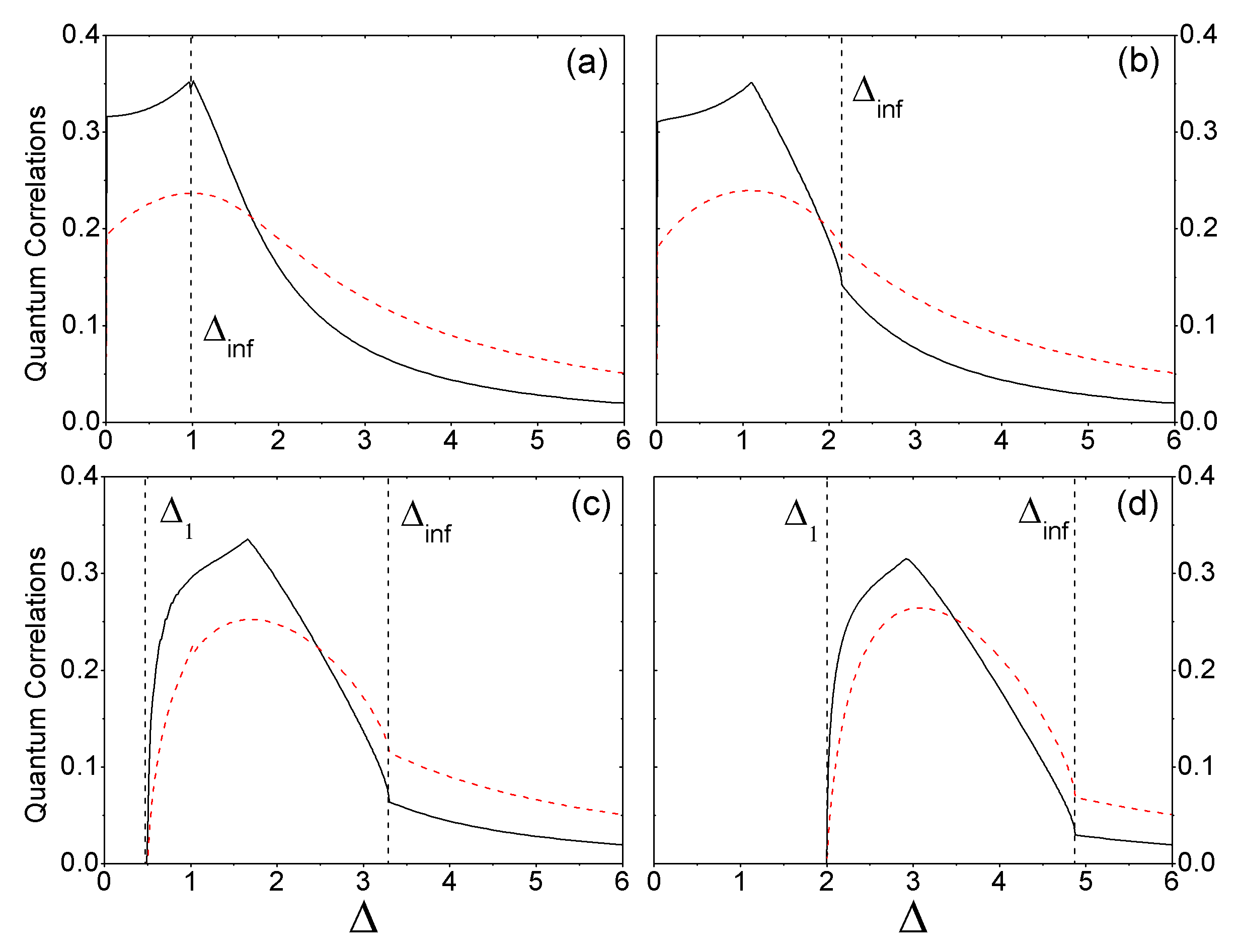}
\caption{\label{fig1}
(color online) $QD$ (black solid line) and EoF (red dashed line) as 
functions of the tuning parameter $\Delta$ and external field $h$ for the XXZ 
model in the thermodynamic limit at $T=0$. (a) $h=0$, (b) $h=2$, (c) $h=6$, 
and (d) $h=12$. The dashed black vertical lines denote the CPs $\Delta_1$ and 
$\Delta_{inf}$. Here and in the following graphics all quantities are 
dimensionless.}
\end{figure}

We begin our analysis by studying the behavior of TQD and EoF as a function of 
the tuning parameter $\Delta$ for $T=0$ and $h\geq0$. These results are shown 
in Fig. \ref{fig1}. For $h=0$ \cite{Dil08,Sar09} 
the infinite-order transition is characterized by 
a global maximum of TQD and EoF at the CP $\Delta_{inf} = 1$ 
(see Fig. \ref{fig1}a). 
But, besides being a maximum, TQD also has a cusp in this CP. This behavior 
implies that the first derivative of TQD is discontinuous at $\Delta_{inf} = 1$ 
while the second derivative is divergent. For the first-order CP 
$\Delta_{1} = -1$ , it was shown in \cite{Sar09} that for $h=0$ the EoF and 
the TQD are discontinuous in the CP.

However, for $h>0$ Figs. \ref{fig1}b-d show that the infinite-order 
transition is characterized through TQD and EoF by a cusp at the CP rather than
a global maximum or minimum. 
For a different spin model it has been recently shown 
\cite{rulli} that the extreme points of entanglement does not necessarily 
indicate an infinite-order transition. Here we show 
an example where an infinite-order CP is pointed out not by a global
 maximum/minimum of the value of a quantum correlation (EoF or TQD). 
Rather, it is a 
discontinuity in its first derivative that spotlight the infinity-order QPT. 
Furthermore, the first-order CP $\Delta_1$ is well determined by TQD and EoF,
as can be seen for $h=6$ (Fig. \ref{fig1}c) and $h=12$ (Fig. \ref{fig1}d). 
Both quantities are zero for $\Delta<\Delta_1$ and nonzero for 
$\Delta>\Delta_1$ while their first derivatives are divergent at $\Delta_1$.
It is worth mentioning that the cusp-like behavior in Fig. \ref{fig1} between
the two CPs is due to an exchange in the set of projectors that minimizes
the quantum conditional entropy \cite{werPRL}, 
Eq. (\ref{conde}), and so far could not be 
associated with any known QPT for this model.

\begin{figure}[!ht]
\includegraphics[width=8cm]{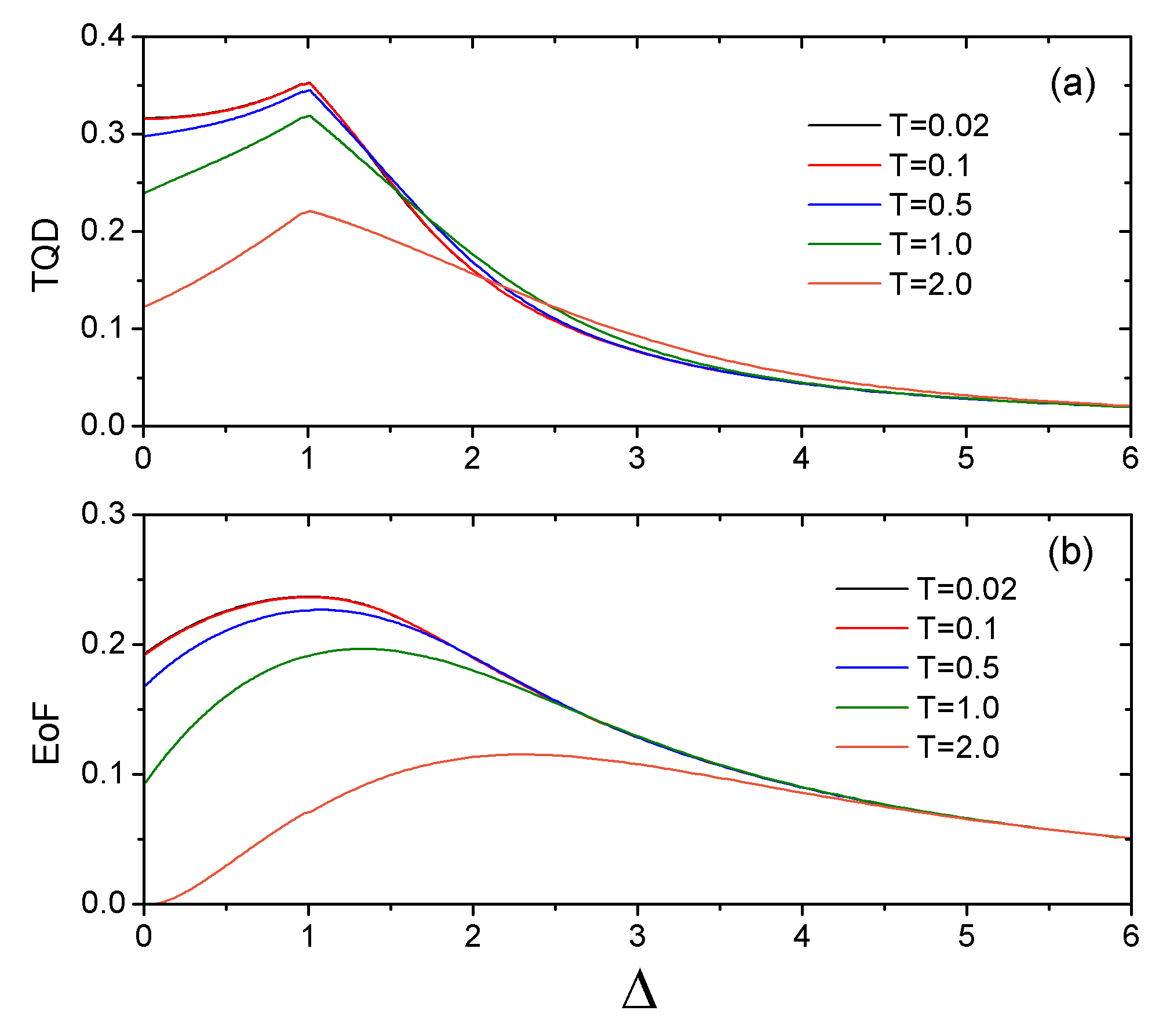}
\caption{\label{fig2} (color online) (a) TQD and (b) EoF as a function of 
$\Delta$ for $h=0$ and, from top to bottom when $\Delta <1$, 
$kT=0.02, 0.1, 0.5, 1.0, 2.0.$ 
The sudden change of TQD remains in the CP $\Delta_{inf}=1$ as the temperature 
increases while the maximum of EoF is shifted to the right.}
\end{figure}

The next step is to examine how the characterization of CPs through TQD and 
EoF is affected by temperature when $h\geq0$. To this end, let us start looking
at the behavior of TQD and EoF for finite $T$ and $h=0$.
In Figs. 
\ref{fig2}a and \ref{fig2}b we plot, respectively, TQD and EoF as a function 
of $\Delta$ for some values of $T>0$ and  $h=0$. As discussed in our 
previous work \cite{werPRL}, and illustrated in 
Fig. \ref{fig2}, for $h=0$ the first-order derivative of TQD is discontinuous 
at the CP $\Delta=1$ not only at $T=0$, but also at $T>0$. On the other hand, 
the EoF is maximum at this CP only when $T=0$ and as $T$ increases the maximum 
moves to higher $\Delta $ values ($\Delta>1$). The behavior of TQD in this case 
is related to the minimization process of the quantum conditional entropy 
(\ref{conde}). We observe that for finite $T$ there is an exchange in the 
functions that minimize the conditional entropy which in turn leads to a 
sudden change in TQD \cite{sudden}, 
characterized by a discontinuity in its first-order derivative. It is worth
noting that for $h=0$ we found a simple closed expression
for TQD involving the two-point correlation functions. This closed expression
together with the fact that at $\Delta =1$ we have 
$\left|\med{\sigma_j^{x}\sigma_{j+1}^{x}}\right|$ $=$ 
$\left|\med{\sigma_j^{z}\sigma_{j+1}^{z}}\right|$
allowed us to explain the sudden change occurring at 
this QPT for $T\geq 0$ \cite{werPRL}. 

\begin{figure}[!ht]
\includegraphics[width=8cm]{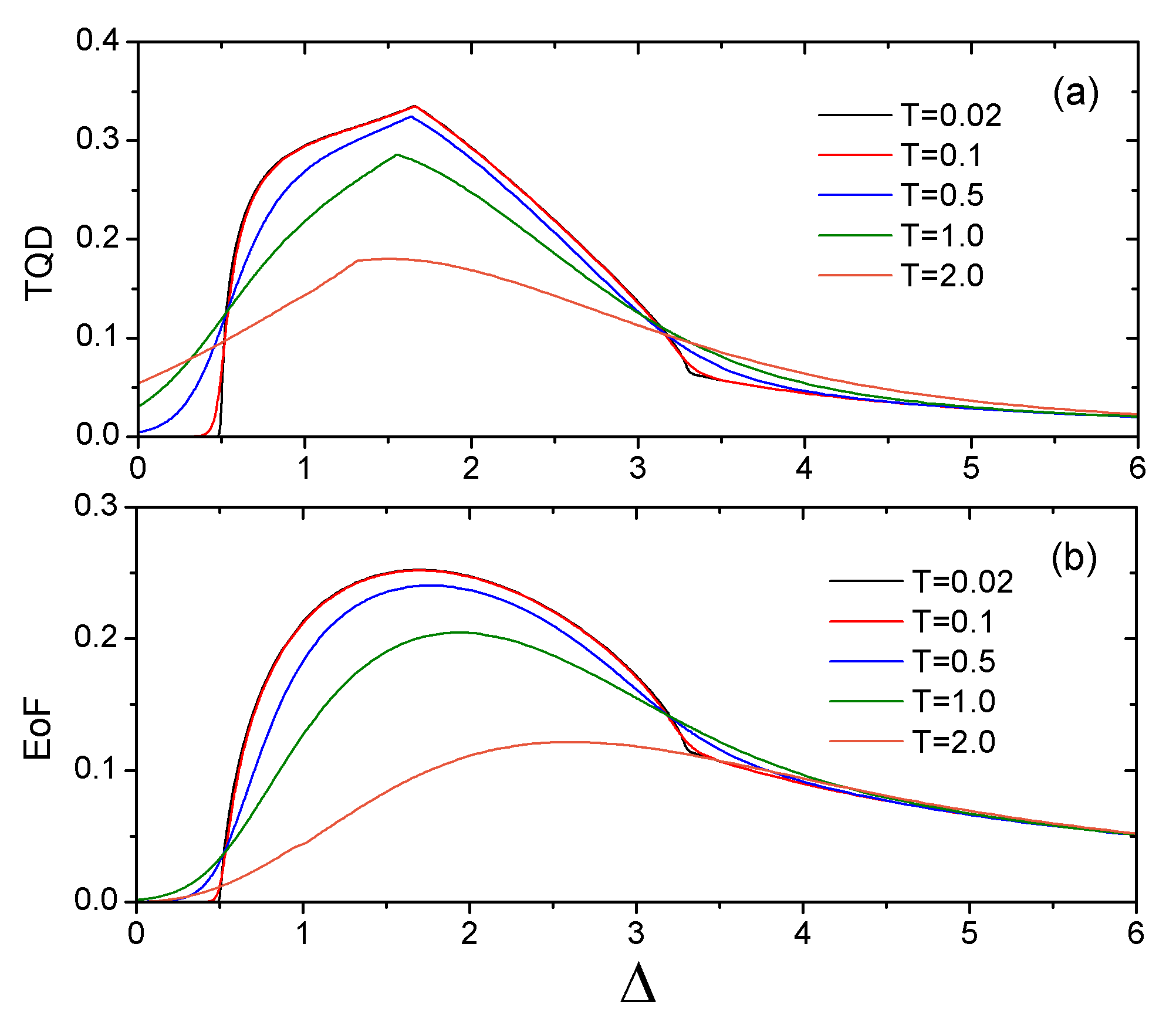}
\caption{\label{fig3} (color online) (a) TQD and (b) EoF as a function of 
$\Delta$ for $h=6$ and, for $\Delta \approx 2$, from top to bottom 
$kT=0.02, 0.1, 0.5, 1.0, 2.0$.}
\end{figure}

\begin{figure}[!ht]
\includegraphics[width=8cm]{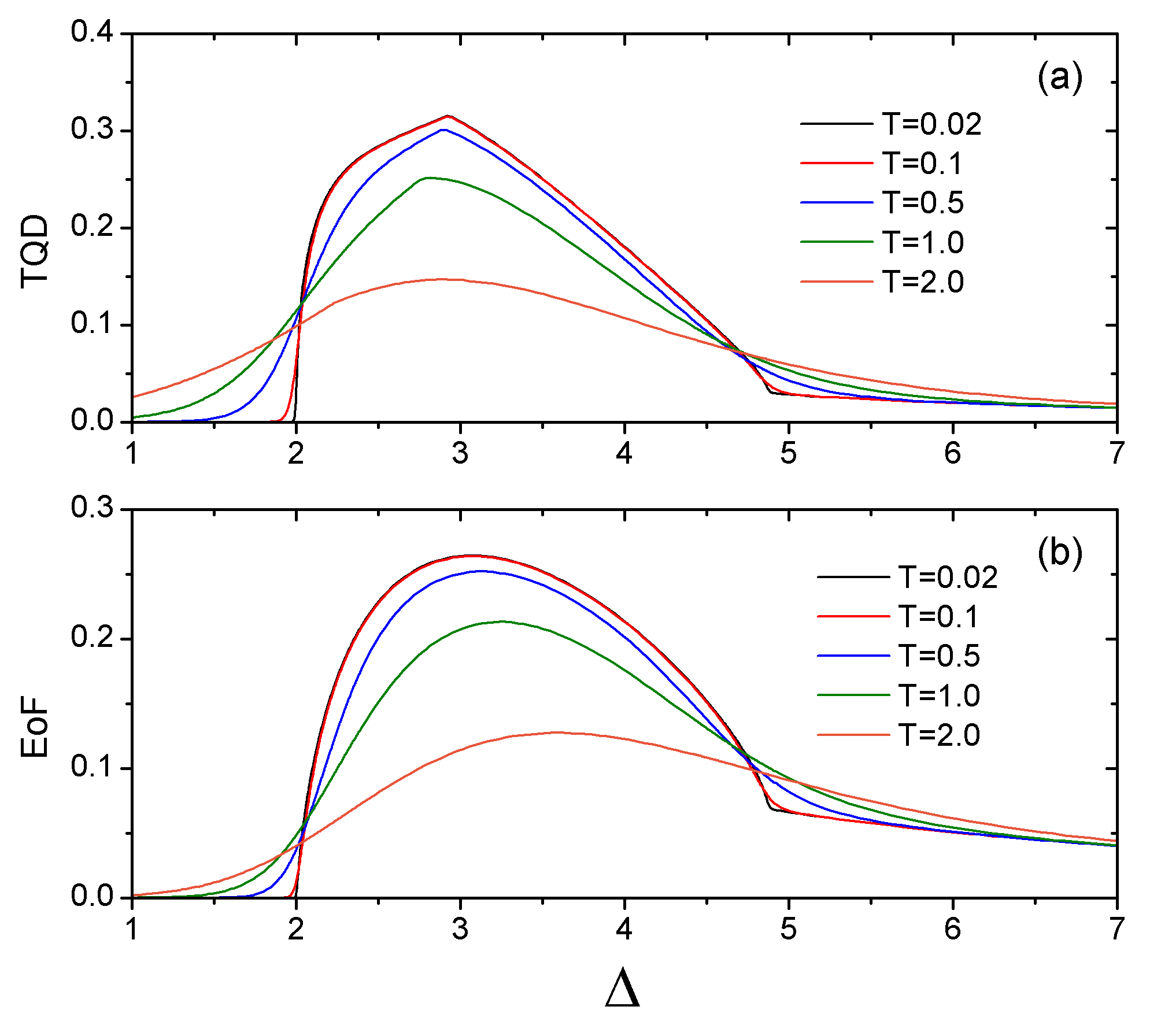}
\caption{\label{fig4} (color online) (a) TQD and (b) EoF as a function of 
$\Delta$ for $h=12$ and, for  $\Delta \approx 3$, from top to bottom 
$kT=0.02, 0.1, 0.5, 1.0, 2.0.$}
\end{figure}

\begin{figure}[!ht]
\includegraphics[width=8cm]{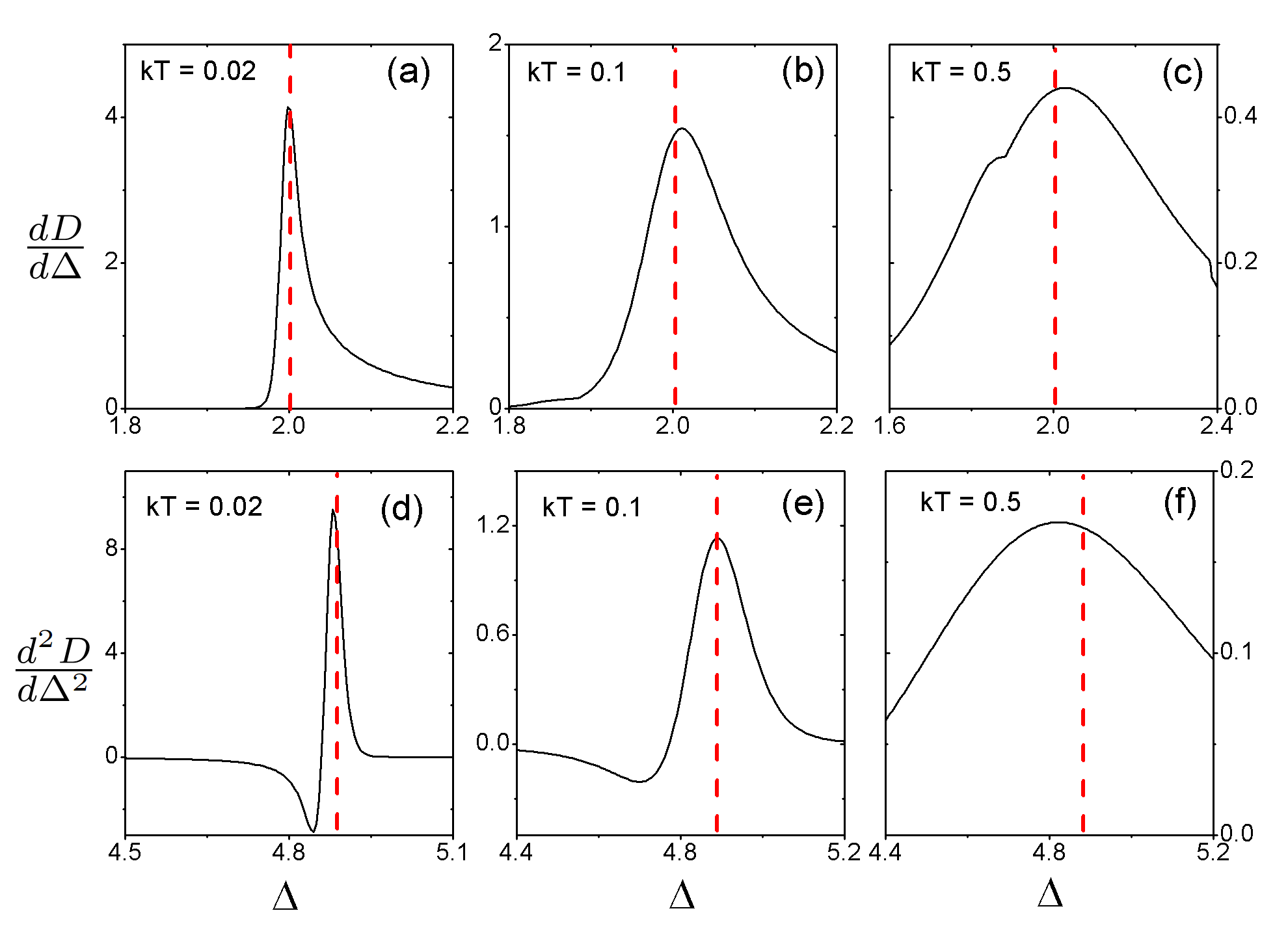}
\caption{\label{fig5} First-order derivative of TQD as a function of $\Delta$ 
for (a) $kT=0.02$, (b) $kT=0.1$, and (c) $kT=0.5$. Second-order derivative of 
TQD as a function of $\Delta$ for (d) $kT=0.02$, (e) $kT=0.1$ and (f) $kT=0.5$.
The maximum of the first and second derivatives of TQD are very close to 
the CPs $\Delta_1=2$ and $\Delta_{inf}=4.88$, respectively. 
Here we fixed $h=12$. Dashed vertical bars indicate the CPs.}
\end{figure}

Moving to the cases where $h>0$ we keep the same set of parameters of 
Fig. \ref{fig2} but the value of the external magnetic field. 
In Figs. \ref{fig3} and \ref{fig4} we use $h=6$ and $h=12$, respectively. 
Looking at Figs. \ref{fig3} and \ref{fig4} we note that as the temperature 
increases the cusp-like behavior of TQD is smoothed while both the maximum of
TQD and EoF decrease. Also, both curves of TQD and EoF become more smooth
and broadened, turning these functions differentiable in the CP.  
Interestingly, however, if the temperature is not too high the 
derivatives of these quantities will still keep some features of their 
behavior at $T=0$. Indeed, as we have shown for $T=0$, the CP $\Delta_1$ is 
characterized by a divergence in the first derivative of both TQD and EoF 
while the CP $\Delta_{inf}$ by a divergence in the second derivative. 
As the temperature increases, though, these divergences disappear but the 
appropriate derivatives of TQD and EoF assume their maximum values around the 
CPs, a property resembling the behavior at $T=0$. To illustrate this point we 
plot in Fig. \ref{fig5} the derivatives of TQD for $h=12$ as a function of 
$\Delta$ and for $kT=0.02, 0.1, 0.5$.
Note that the first-order derivative of TQD has a local maximum near the CP 
$\Delta_1=2$ for $kT=0.02, 0.1, 0.5$. The infinite-order QPT is also well 
characterized by a local maximum in the second-order derivative of TQD near 
the CP $\Delta_{inf}=4.88$ for $kT=0.02, 0.1, 0.5$. 
This procedure to determine, or at least estimate with great
accuracy, the CPs with finite 
temperature theoretically computed data has been tested for 
other values of $h$ and it was proved valid. The same analysis can be applied 
to EoF but, as we show in the following, TQD gives better results than EoF for 
all sets of parameters.

\begin{figure}[!ht]
\includegraphics[width=8cm]{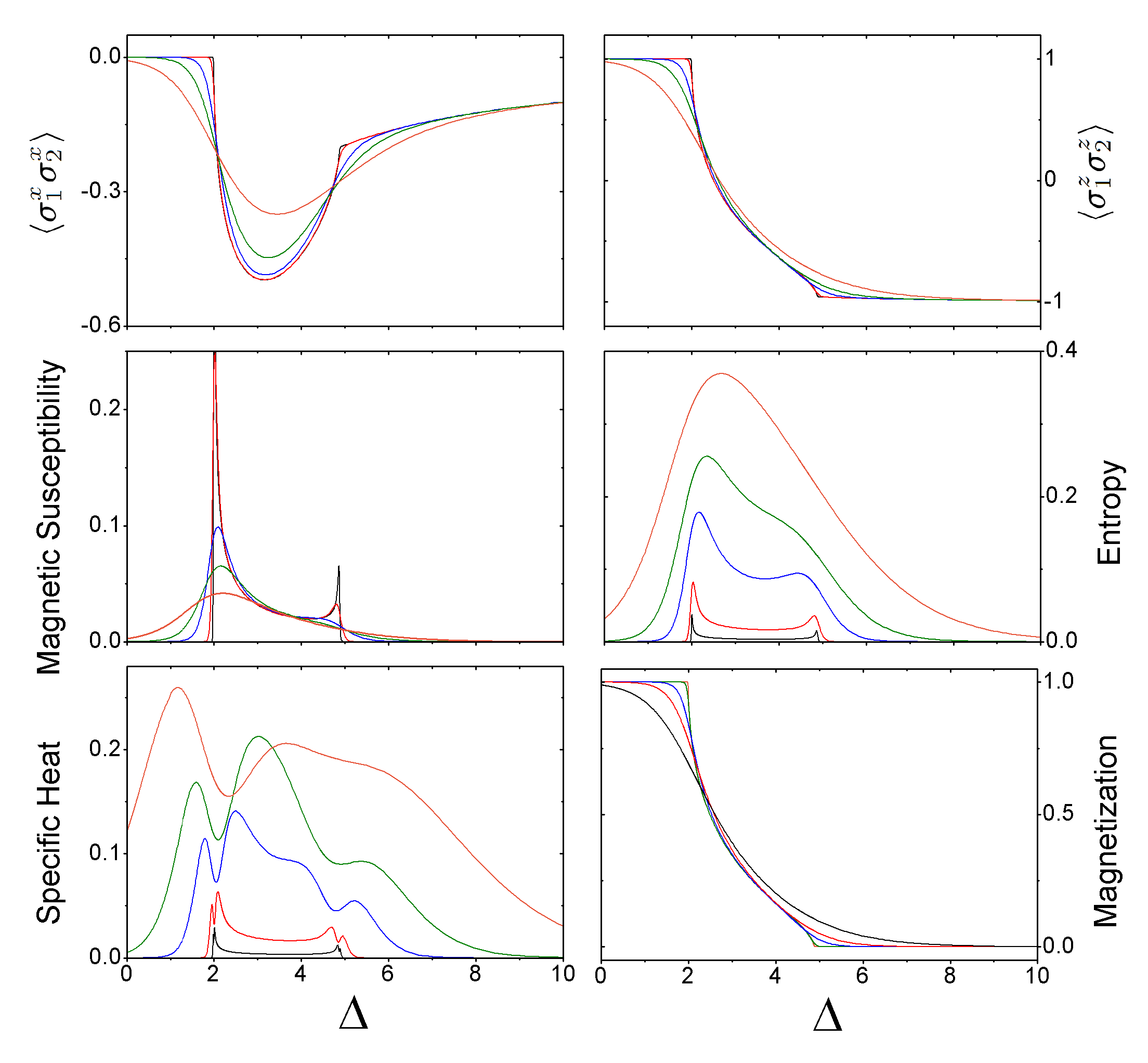}
\caption{\label{fig6} (color online) Thermodynamic quantities for the XXZ 
model in the thermodynamic limit for $h=12$ and $kT=0.02$ (black line), 
$kT=0.1$ (red line), $kT=0.5$ (blue line), $kT=1.0$ (green line), and 
$kT=2.0$ (orange line). Here the CPs are 
$\Delta_1=2$ and $\Delta_{inf}=4.88$. For $\Delta < 1$ and for 
the curves of the magnetization and the two-point correlations $kT$
increases from top to bottom while the opposite happens to the curves
of the other quantities.}
\end{figure}

\begin{figure}[!ht]
\includegraphics[width=8cm]{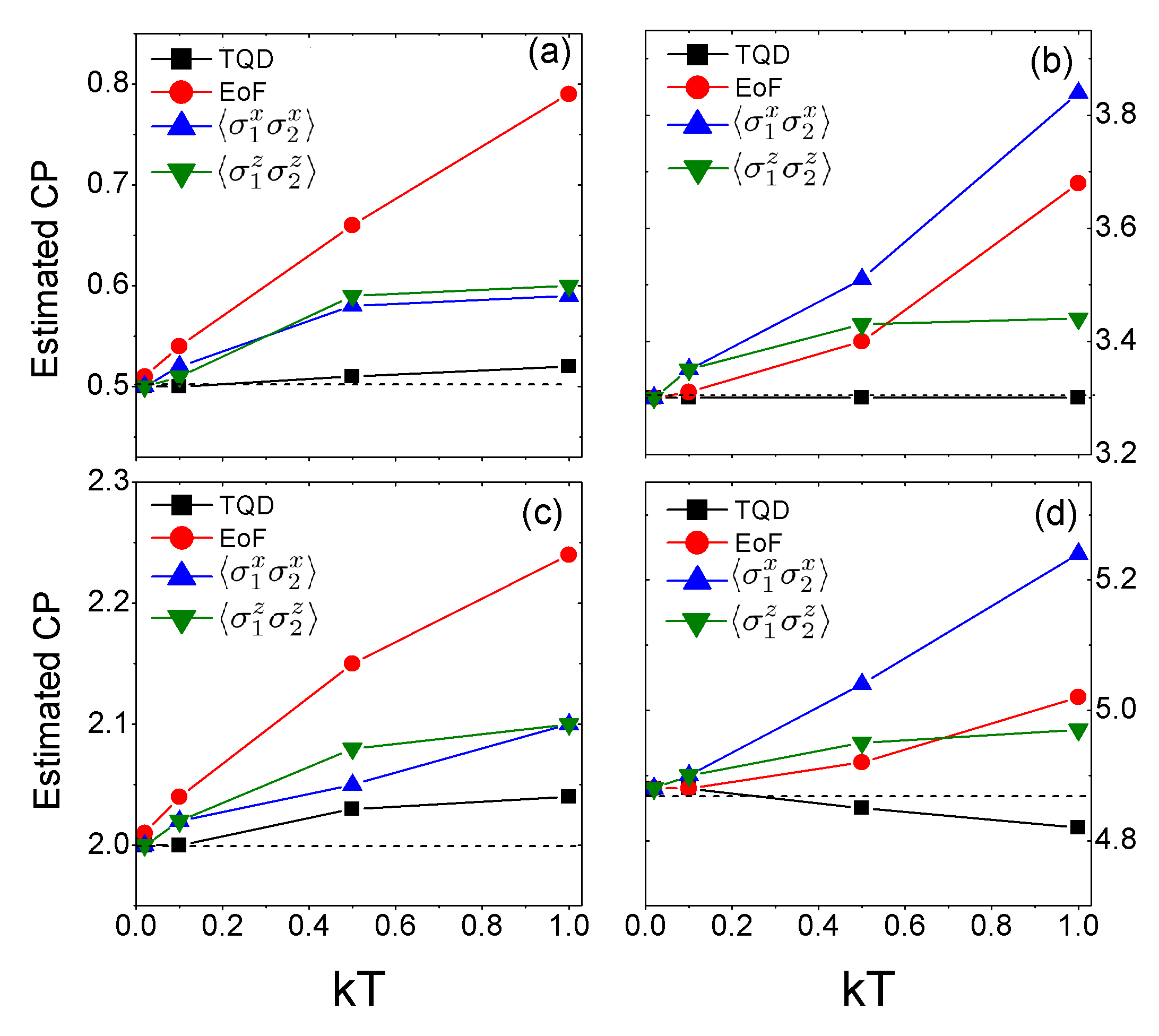}
\caption{\label{fig7} (color online) CPs determined by TQD (square), EoF 
(circle), $\med{\sigma_1^x\sigma_2^x}$ (up arrow), and 
$\med{\sigma_1^z\sigma_2^z}$ (down arrow) as a function of $T$. 
In (a) and (b) we have $h=6$ with a first (a) and an infinite-order (b) CP; 
in (c) and (d) we have $h=12$ with a first (c) and an infinite-order (d) CP. 
Dashed horizontal lines indicate the CPs. To determine the CPs at 
finite $T$ we used the derivatives of TQD and EoF. As stated in the text, 
for some temperature ranges, the derivatives of these quantities at finite 
$T$ retain some information about the divergences or discontinuities that 
exist only at $T=0$. Note that the estimated CPs given by TQD is always 
closer to the exact value than the CPs coming from other quantities. }
\end{figure}

In addition, in Fig. \ref{fig6} we show, for 
several temperatures, many thermodynamic quantities for the infinite spin 
chain and also the pairwise correlations as a function of $\Delta$ for $h=12$.
For low temperatures the CPs can be estimated by some thermodynamic 
quantities such as the magnetic susceptibility, entropy, and specific heat. 
Although TQD is not the only quantity that can determine or estimate 
a CP close to $T=0$ (for $h>0$) it is the best {\it CP detector} compared with 
other quantities tested here. This is shown in Fig. \ref{fig7}, 
where we plotted the CPs determined via TQD, EoF, and the pairwise 
correlations  $\left\langle \sigma^x_1\sigma^x_2\right\rangle$ and 
$\left\langle \sigma^z_1\sigma^z_2\right\rangle$ for several temperatures.
We chose to work with $h=6$ and $h=12$ but similar results are obtained for 
other values of the magnetic field. 

The procedure adopted to estimate the CPs when $T$ is finite comes from 
the behavior of TQD,  EoF, and the two point correlations 
near or at the CPs. For example, in Figs. 
\ref{fig2}, \ref{fig3}, and \ref{fig6} we see that in the vicinity of the CPs 
these quantities, as a function of $\Delta$, are either maximum or possess 
discontinuous/divergent first/second derivatives. Therefore, by computing
the appropriate derivative and looking for its maximum/minimum we can estimate
the value of the CPs even for finite $T$. Obviously, for $T=0$ these 
extrema are exactly at the CP.  As can be seen in Fig. \ref{fig7}, 
the CPs estimated by TQD are closer to the values given in Table I than 
those obtained using EoF or the pairwise correlations. Indeed, from 
zero to $kT \approx 1.0$ the extrema of TQD is closer to the actual CP than 
those from the other quantities (see Fig. \ref{fig7}).

\begin{figure}[!ht]
\includegraphics[width=8cm]{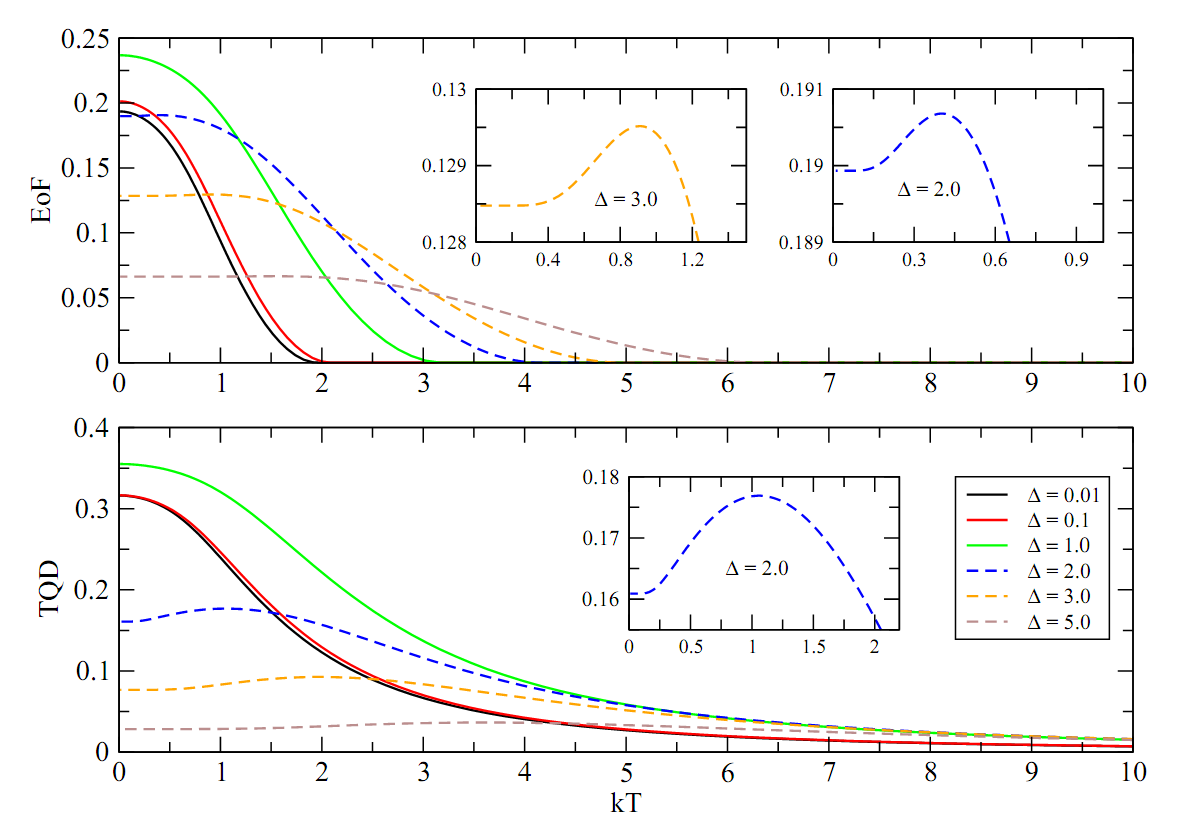}
\caption{\label{fig8} (color online) TQD (bottom) and EoF (top) as a function 
of $T$ for different values of the tuning parameter $\Delta$. The insets show 
the regrowth of TQD and EoF. For $\Delta \leq 1$ (solid curves) we note that 
as we increase $\Delta$ we also increase TQD and EoF. 
For $\Delta > 1$ (dashed curves)
this only happens for EoF at high temperatures.}
\end{figure}

To complement our studies about the XXZ model we present in Fig. \ref{fig8} 
the behavior of TQD and EoF as a function of temperature for $h=0$. 
First, we observe that EoF tends to disappear suddenly as the temperature 
increases while TQD is always nonzero. This sudden disappearance may be 
retarded by increasing the value of the anisotropy $\Delta$. The insets in 
Fig. \ref{fig8} show a ``regrowth'' \cite{Wer10} of TQD and EoF with 
temperature. This interesting feature 
was already seen for the behavior of TQD of a two-spin XXZ model with no field  
\cite{Wer10} although the regrowth does not show up for EoF \cite{Rig04}. 
Therefore, the results here presented show that the regrowth of EoF 
with temperature, while not observed for two-spin chains without the presence 
of an external field \cite{Rig04}, 
is possible in the thermodynamic limit.

\subsection{The XY Model}

In this section we discuss the ability of TQD and EoF to point out CPs of 
a QPT in the infinite one-dimensional XY model in a transverse field
for finite $T$. 
The Hamiltonian that describes this model is given by 
\begin{eqnarray}
\label{HXY}
H_{xy}&=& - \frac{\lambda}{2} \sum_{j=1}^L \left[(1+\gamma)\sigma^x_j
\sigma^x_{j+1}+(1-\gamma)\sigma^y_j\sigma^y_{j+1}\right] \nonumber \\
&&  -\sum_{j=1}^L\sigma^z_j.
\end{eqnarray}
The strength of the inverse of the external transverse magnetic field is 
represented by $\lambda$ while $\gamma$ provides the degree of anisotropy. 
For $\gamma=\pm1$ we obtain the transverse Ising model while $\gamma=0$ 
corresponds to the XX model in a transverse field \cite{LSM}, 
which is the $\Delta=0$ XXZ model discussed before with $J<0$. 
As is well-known the 
XY model undergoes a second-order QPT (Ising transition \cite{isingQPT}) 
at the CP $\lambda_c=1$ \cite{Dil08}, that separates a ferromagnetic ordered 
phase from a quantum paramagnetic phase. For $\lambda>1$ one can 
observe another second-order QPT at the CP $\gamma_c=0$ called anisotropy 
transition \cite{LSM,anisQPT}. 
Differently from the Ising transition, that is due 
to the action of the external field, this transition is driven by the 
anisotropy parameter $\gamma$ and separates a ferromagnet ordered along 
the $x$ direction and a ferromagnet ordered along the $y$ direction. 
Although both QPT have the same order they belong to different universality 
classes \cite{LSM,anisQPT}.

The XY Hamiltonian is $Z_2$-symmetric and can be exactly diagonalized 
\cite{LSM} in the thermodynamic limit $L\rightarrow\infty$. 
Due to the translational invariance the two-spin density operator 
$\rho_{i,j}$ for spins $i$ and $j$ depends only on the distance  
$k=|j-i|$ between them and thus $\rho_{i,j}=\rho_{0,k}$. Therefore, 
the reduced density operator $\rho_{0,k}$ for the XY model at thermal 
equilibrium is \cite{osborne}
\begin{eqnarray}\label{doXY}
\rho_{0,k}&=&\frac{1}{4}\left[I_{0,k}+
\left\langle \sigma^z\right\rangle\left(\sigma^z_0+\sigma^z_k\right)\right] 
\nonumber \\
&& + \frac{1}{4} \sum_{\alpha=x,y,z} 
\left\langle \sigma^\alpha_0\sigma^\alpha_k\right\rangle 
\sigma^\alpha_0\sigma^\alpha_k,
\end{eqnarray}
where $I_{0,k}$ is the identity operator of dimension four.  
The transverse magnetization $\left\langle \sigma^z_k\right\rangle$
$=$ $\left\langle \sigma^z\right\rangle$ is
\begin{eqnarray}\label{tmag}
\left\langle \sigma^z\right\rangle=-\int_0^\pi 
(1+\lambda\cos{\phi})\tanh{(\beta\omega_\phi)}\frac{d\phi}{2\pi\omega_\phi},
\end{eqnarray}
with $\omega_\phi=\sqrt{(\gamma\lambda\sin{\phi})^2+(1+\lambda\cos{\phi})^2}/2$. 
The two-point correlation functions are given by
\begin{eqnarray}\label{tpcf}
\left\langle \sigma^x_0\sigma^x_k\right\rangle &=& \left|
\begin{array}{cccc}
G_{-1} & G_{-2} & \cdots & G_{-k}\\
G_{0} & G_{-1}  & \cdots & G_{-k+1} \\
\vdots & \vdots & \ddots & \vdots \\
G_{k-2} & G_{k-3} & \cdots & G_{-1}  \\
\end{array}
\right|,\\
\left\langle \sigma^y_0\sigma^y_k\right\rangle &=& \left|
\begin{array}{cccc}
G_{1} & G_{0} & \cdots & G_{-k+2}\\
G_{2} & G_{1}  & \cdots & G_{-k+3} \\
\vdots & \vdots & \ddots & \vdots \\
G_{k} & G_{k-1} & \cdots & G_{1}  \\
\end{array}
\right|,\\
\left\langle \sigma^z_0\sigma^z_k\right\rangle &=& 
\left\langle \sigma^z\right\rangle^2 - G_k G_{-k},
\end{eqnarray}
where
\begin{eqnarray*}
G_k&=&\int_0^\pi \tanh{(\beta\omega_\phi)}\cos{(k\phi)}(1+\lambda\cos{\phi}) 
\frac{d\phi}{2\pi\omega_\phi}\\
&-&\gamma\lambda\int_0^\pi \tanh{(\beta\omega_\phi)} \sin{(k\phi)\sin{\phi}} 
\frac{d\phi}{2\pi\omega_\phi}.
\end{eqnarray*}  
Using these relations and Eqs. (\ref{qdX}) and (\ref{eof}) of the Appendix, 
EoF and TQD 
can be computed analytically for this model. 

The ability of TQD and EoF to detect a CP for this model 
at $T=0$ was studied 
in \cite{Sar10}, where TQD and EoF from first to fourth 
nearest-neighbors was computed. 
They showed that TQD between far neighbors can still characterize a QPT while 
EoF is zero. This result shows an advantage of the quantum discord to the 
entanglement in the detection of CP. See also \cite{Ami10} for the behavior 
of entanglement and TQD for small $T$ in the symmetry breaking process of the
XY model. Our goal here is to determine whether the 
quantum discord for the first and second neighbors has some advantage with 
respect to entanglement to spotlight a CP at finite $T$. The procedure adopted
here is similar to the one employed in the analysis of the XXZ model, i.e.,
we compute the appropriate derivatives around the CP and look for its 
extremum values as indicators of QPTs. 

\begin{figure}[!ht]
\includegraphics[width=8cm]{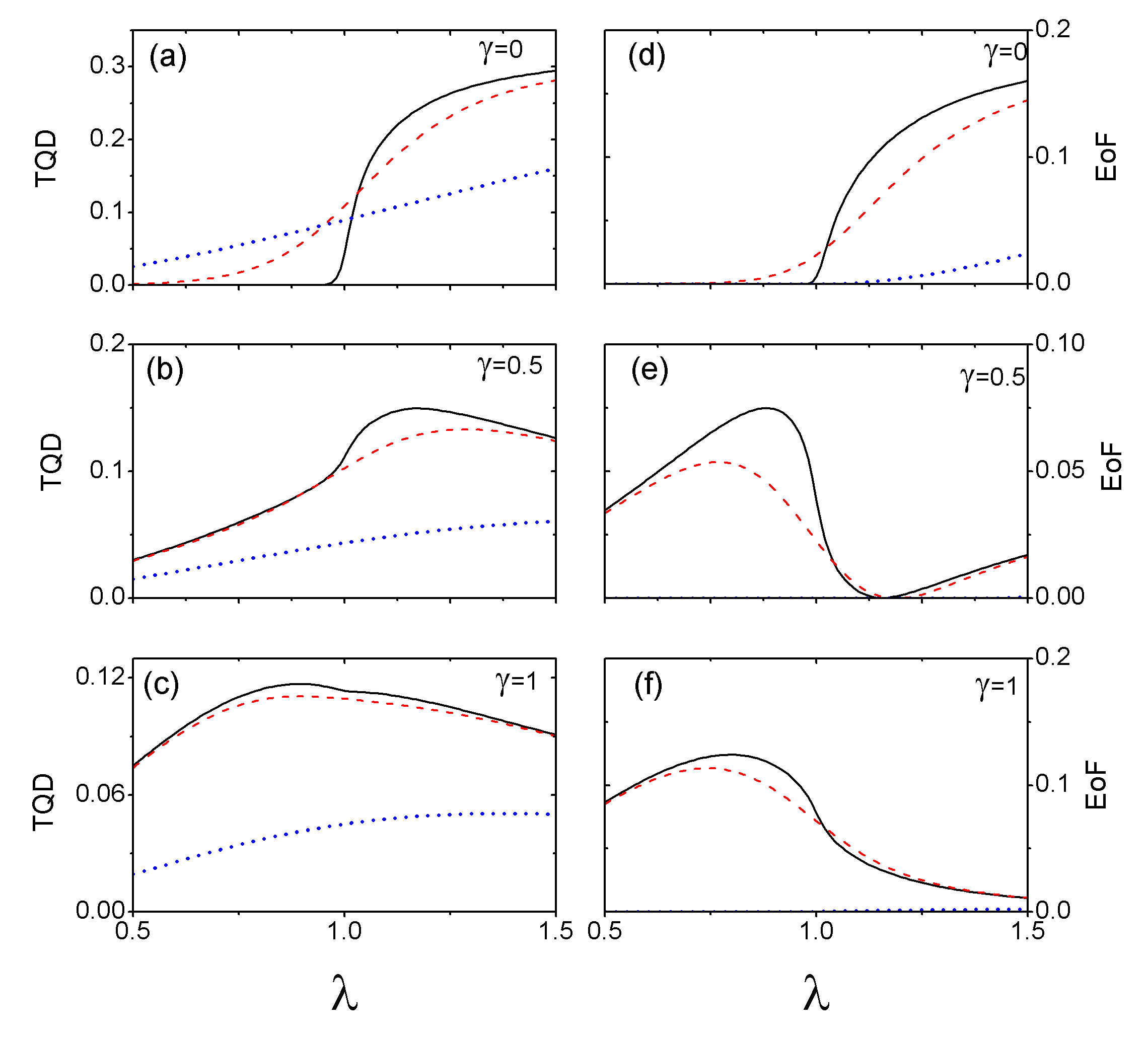}
\caption{\label{fig9} (color online) (a)-(c) TQD and (d)-(f) EoF as a function 
of $\lambda$ for $kT=0.01$ (black/solid line), $kT=0.1$ (red/dashed line) and 
$kT=0.5$ (blue/dotted line) for nearest-neighbors. 
We use three values of $\gamma$ as shown in the graphs.}
\end{figure}

\begin{figure}[!ht]
\includegraphics[width=8cm]{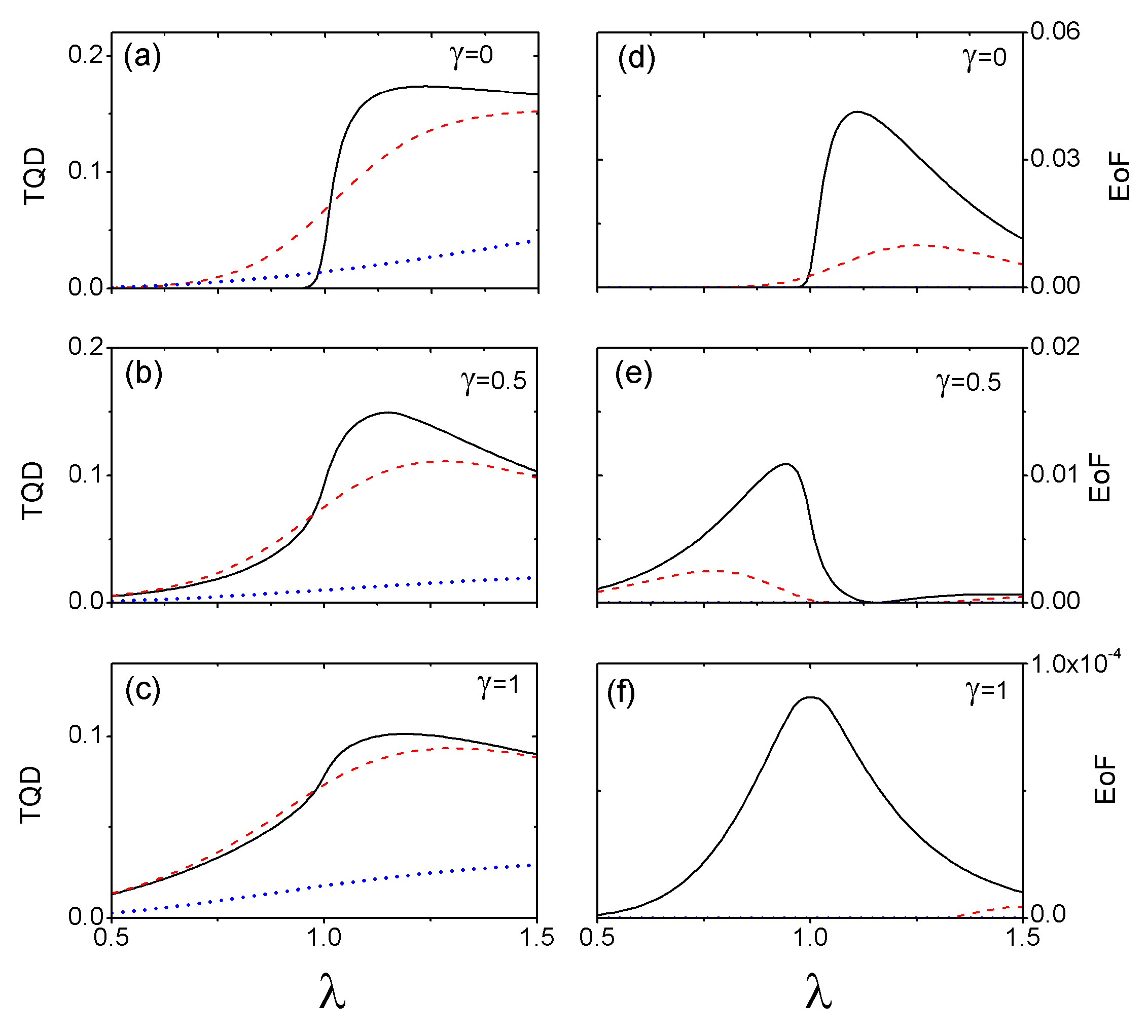}
\caption{\label{fig10} (color online) (a)-(c) TQD and (d)-(f) EoF as a function
of $\lambda$ for $kT=0.01$ (black/solid line), $kT=0.1$ (red/dashed line) and 
$kT=0.5$ (blue/dotted line) for second nearest-neighbors. We use three values 
of $\gamma$ as shown in the graphs.}
\end{figure}

In Figs. \ref{fig9} and \ref{fig10} we plot, respectively,  
TQD and EoF for first nearest-neighbors and second nearest-neighbors as a 
function of $\lambda$ for $kT=0.01, 0.1, 0.5$ and $\gamma=0, 0.5, 1.0$.  
It is important to note that TQD is more resistant to thermal effects than 
EoF. For example, for $kT=0.5$ EoF for the first-neighbors is zero or close to 
zero for almost all $\lambda$ while TQD is always non-null. For 
second-neighbors the situation is more drastic,  EoF is always zero for 
$kT=0.5$. The previous discussion also highlights a qualitative difference 
between the behavior of quantum correlations as measured by quantum discord 
or entanglement, especially for $\gamma=0.5$.

\begin{figure}[!ht]
\includegraphics[width=8cm]{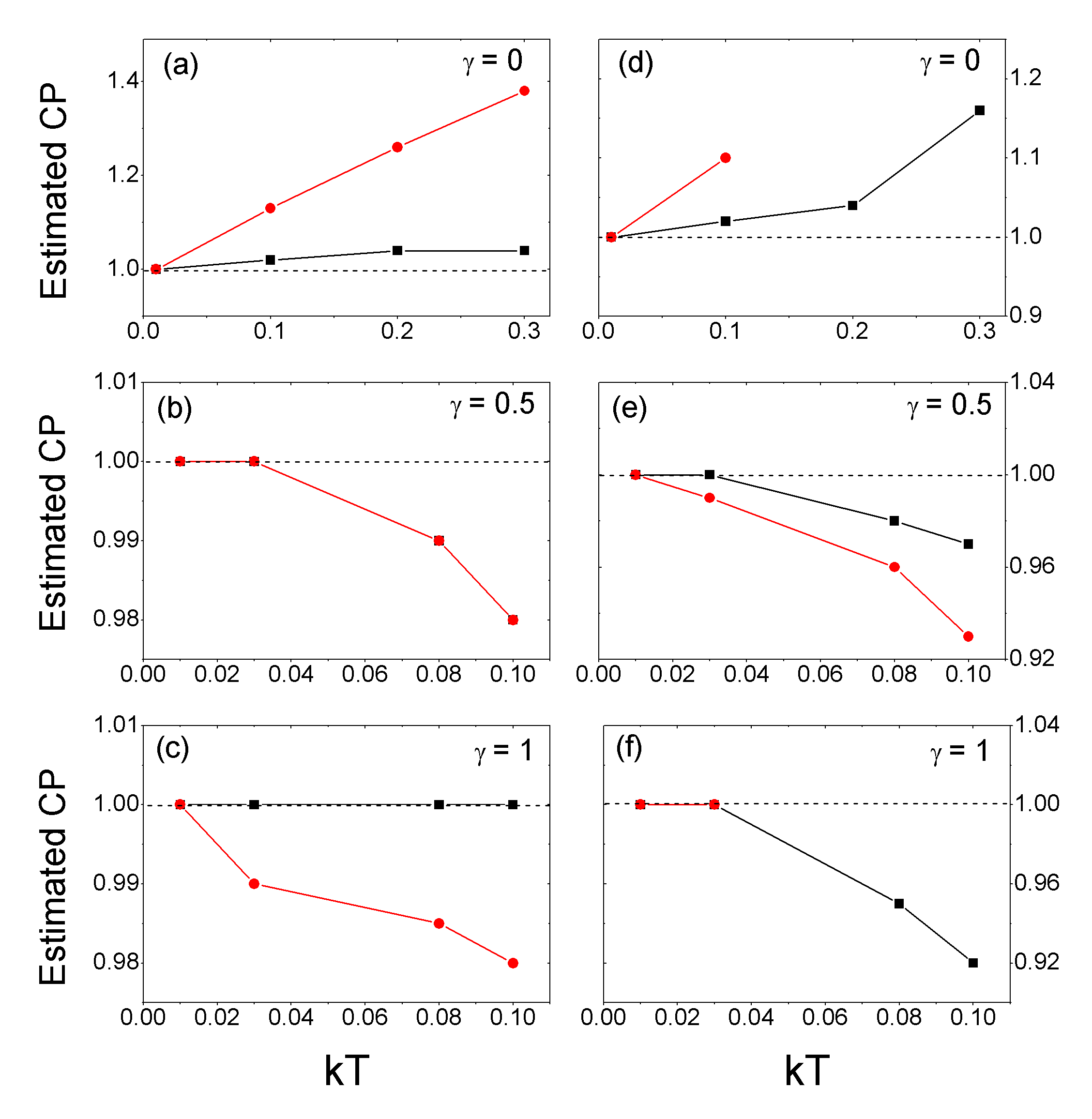}
\caption{\label{fig11} (color online)  CPs determined by TQD (black-squares) 
and EoF (red-circles) as a function of $kT$ for (a)-(c) first nearest-neighbors
and for (d)-(f) second nearest-neighbors. 
Dashed horizontal lines indicate the CPs. The values of $\gamma$ used were 
$0$, $0.5$ and $1.0$, as shown in the graphs. The derivatives of TQD and EoF 
were used to estimate the CPs as explained in the text. 
Note that at panel (b) the curves for TQD and EoF coincide and that 
at panels (d) and (f) we only have two points for EoF since 
for greater $T$ it is zero.}
\end{figure}

Since the Ising transition is a second-order QPT, 
the critical point at $T=0$ is characterized by a divergence or discontinuity 
in the first derivative of TQD or EoF \cite{Dil08,Sar09}; and if the first 
derivative is discontinuous then the CP can also be found through the 
divergence of the second derivative. Similarly to the XXZ model studied
in Sec. \ref{secXXZ}, the singular behavior of TQD and EoF at the CP 
is attenuated as the temperature increases. However, one can still extract 
useful information concerning the CPs until a certain temperature. 
Beyond this temperature the behavior of the derivatives of TQD or EoF is not 
able to determine the CPs without ambiguity. With that in mind,
we use the same strategy of Sec. \ref{secXXZ} to determine/estimate the CPs 
with theoretically computed data at finite $T$: 
if the first derivative of TQD or EoF is 
divergent at $T=0$ then the CP is pointed out by a local maximum or minimum at 
$T>0$; if the first derivative is discontinuous at $T=0$ then we look
after local maximum or minimum in the second derivative for $T>0$. 
The results of our analysis are shown in Fig. \ref{fig11} 
where the CPs estimated by TQD and EoF for first and second neighbors are 
plotted as a function of $kT$ for $\gamma=0, 0.5$, and $1.0$. 

Looking at Fig. \ref{fig11} we note that the CP of the Ising transition is 
best estimated by TQD than EoF. 
While for $\gamma=0.5$ and $\gamma=1$ the first-neighbor TQD is only slightly 
better than EoF to estimate the CP , for 
$\gamma=0$ the CP computed by TQD is clearly closer to $\lambda_c=1$ than the
one estimated from EoF. Moreover, the temperature below which TQD is a good 
CP detector is higher for $\gamma=0$ than for $\gamma=0.5$ or $\gamma=1$ 
(see Fig. \ref{fig11}a). When we look at the second-neighbors TQD and EoF,
we see that TQD is still  
better than EoF to estimate the CP. And in particular
for $\gamma=1.0$,  EoF is only useful for very tiny $T$ while TQD works
relatively well for higher values of $T$. 
It is worth noting that for small $T$ TQD seems to approach 
the actual CP almost linearly as we decrease $T$. 
Therefore, a relatively small number of points for a couple
of small temperatures allows one to correctly extrapolate to the exact value of 
the CP at $T=0$. 
This fact is also observed for the behavior of TQD in the XXZ model. An
interesting future investigation, lying beyond the scope of this 
manuscript, would be to rigorously determine the
functional form by which TQD and EoF approach the exact CP for
decreasing $T$.

\begin{figure}[!ht]
\includegraphics[width=8cm]{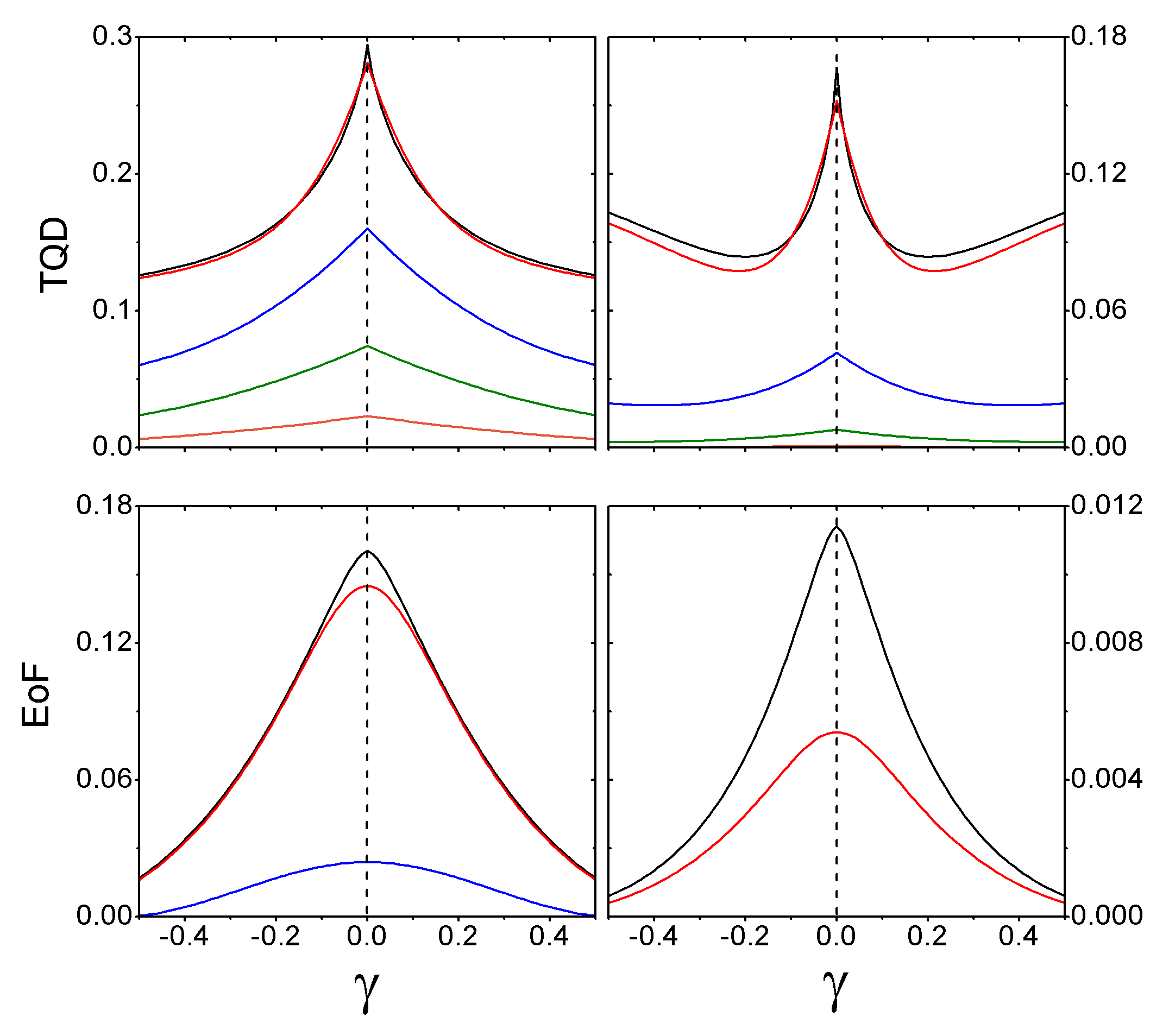}
\caption{\label{fig12} (color online) Top panels TQD and bottom
panels EoF as a 
function of $\gamma$. From top to bottom $kT=0.001, 0.1, 0.5, 1.0$, and
$2.0$.  Left panels are first nearest-neighbors and right
panels second nearest-neighbors. Here we fixed $\lambda = 1.5$
and the CP is $\gamma_c=0$ (indicated by dashed
vertical lines).}
\end{figure}

We close this section studying the behavior of  TQD and EoF near the critical 
point $\gamma_c=0$, which is associated to the anisotropy transition that 
occurs only for $\lambda>1$. Hence, for definiteness, in what follows we set 
$\lambda=1.5$.  In Fig. \ref{fig12} we plot TQD and EoF for the first and 
second-neighbors as a function of $\gamma$ for 
five different values of 
temperature,  $kT=0.001, 0.1, 0.5, 1.0$, and $2.0$. 

Looking at Fig. \ref{fig12} we see that 
TQD and EoF are maximum at the CP  $\gamma_c=0$ for 
first- and second-neighbors, 
keeping this feature as $T$ increases. However, only TQD possesses
a cusp-like behavior at the CP. Furthermore, for first-neighbors 
TQD presents the same pattern (maximum with a cusp-like behavior)
up to  $kT=2.0$ while EoF vanishes for 
$kT\geq1.0$. This situation is more drastic for second-neighbors: 
while TQD can detect the CP even for values near $kT=1.0$, EoF is nonzero only 
at low temperatures ($kT\lesssim0.1$). Finally, as in the XXZ model with
no external field, there is an exchange in the set of projectors that minimizes 
the conditional entropy (\ref{conde}) at the CP. This feature is 
characterized by a discontinuity in the first-order derivative of TQD and 
explains the cusp-like behavior of TQD at $\gamma_c=0$.

\section{Conclusions}

In this paper we have studied the ability of quantum correlations to
spotlight critical points of quantum phase transitions when the system 
under investigation is away from the absolute zero, the experimentally 
unattainable temperature where a quantum phase transition sets in. We 
have mainly focused our analysis in two distinct types of quantum correlations,
namely, the entanglement and the quantum discord between pairs of spins. 

We first concentrated our efforts in computing those quantum correlations
for an infinite chain described by the XXZ model with an external field in 
the $z$-direction and in equilibrium with a thermal reservoir at temperature
$T$. By tracing out all but two nearest-neighbors we were able to calculate
their entanglement and quantum discord for several T. For this model we 
also evaluated several thermodynamic quantities as well. Among all these
quantities, we have shown that quantum discord is the best critical 
point detector at finite $T$, furnishing rather precise values for 
the two quantum critical points for this model, a first order and an
infinite order phase transition.

We then moved to study other models, such as the infinite chain XY model and
the Ising model, both of which with external transverse magnetic fields. 
This time we computed the entanglement and the quantum discord for first 
and second nearest-neighbors, assuming the system in equilibrium with a 
thermal reservoir at temperature $T$. For the second order phase transition
usually referred to as the Ising transition, we observed that quantum discord
is better suited in estimating the correct value for the quantum critical
point. For the other quantum phase transition, the so called 
anisotropy transition (also second order), quantum discord and entanglement
give excellent results for low temperatures. However, as the temperature
increases quantum discord outperform entanglement as a quantum critical point 
estimator.  

Summarizing, the previous discussions showed that 
for all the cases here presented, quantum discord was the best critical point 
estimator when one deals with finite temperature systems. 
Note that we never needed to know
the order parameter of the phase transitions to use these quantum correlations
as critical point detectors. Moreover, we have also showed
that using the quantum discord (our best critical point estimator), 
and with theoretically computed data for several $T$, we can with 
reasonable accuracy 
infer the actual value of the critical point at $T=0$ by looking at the 
behavior of these quantities when $T$ approaches zero. Therefore, 
we have given strong indications that entanglement and in special 
quantum discord are useful quantities in estimating quantum critical points 
when the available data are away from the absolute zero. This fact alone 
shows that these quantities 
are an important tool in the study of quantum phase transitions in realistic
experimental scenarios, where one is always working at finite $T$ and has 
perhaps no clue of the order parameter of the phase transition taking place.

\begin{acknowledgments}
TW and GR thank CNPq for funding. GAPR thanks CNPq/FAPESP for funding.
GR also thanks CNPq/FAPESP for financial support through the
National Institute of Science and Technology for Quantum Information.
\end{acknowledgments}

\appendix* 

\section{Closed expressions for QD and EoF}

In Sec. \ref{results}, 
after tracing out all spins but two in 
an infinite chain,
the two-qubit reduced density matrix is of X-form,
\begin{equation}\label{doX}
\rho_{AB}=\left(\begin{array}{cccc}
\rho_{11} & 0 & 0 & \rho_{14}\\
0 & \rho_{22} & \rho_{23} & 0\\
0 & \rho_{23} & \rho_{22} & 0\\
\rho_{14} & 0 & 0 & \rho_{44}\
\end{array}\right).
\end{equation}
It is easy to see that for this density matrix 
$\mathcal{S}(A)=\mathcal{S}(B)=-\sum_{i=1}^2\beta_i\log_2\beta_i$, with 
$\beta_1=\rho_{11}+\rho_{22}$ and $\beta_2=\rho_{44}+\rho_{22}$, implying that 
QD is a 
symmetric quantity. For the density operator (\ref{doX}) this minimization 
can be done analytically, as previously demonstrated in \cite{discordX}, 
resulting in the following 
expression,
\begin{eqnarray}\label{conde}
\mathcal{S}_q(A|B)\!=\!\min\!\left\{ \beta_1\mathcal{F}(\theta_1) + 
\beta_2\mathcal{F}(\theta_2), \mathcal{F}(\theta_3),\mathcal{F}(\theta_4)  
\right\},\nonumber \\
\end{eqnarray}
where $\mathcal{F}(\theta)=-\frac{1-\theta}{2}\log_2\frac{1-\theta}{2}-
\frac{1+\theta}{2}\log_2\frac{1+\theta}{2}$ and 
\begin{eqnarray*}
\theta_1&=&\frac{1}{\beta_1}\left|\rho_{22}-\rho_{11}\right|,\\
\theta_2&=&\frac{1}{\beta_2}\left|\rho_{22}-\rho_{44}\right|,\\
\theta_3&=&\sqrt{4(\rho_{14}-\rho_{23})^2+(\rho_{11}-\rho_{44})^2},\\
\theta_4&=&\sqrt{4(\rho_{14}+\rho_{23})^2+(\rho_{11}-\rho_{44})^2}.
\end{eqnarray*}
Using (\ref{conde}) in Eq. (\ref{qd}) and calculating the joint 
entropy $\mathcal{S}(A,B)$ of the density operator (\ref{doX}),
the quantum discord is obtained as 
\begin{eqnarray}\label{qdX}
D\left(\rho_{AB}\right)&=&-\sum_{i=1}^2\beta_i\log_2\beta_i +\sum_{j=1}^4
\lambda_j\log_2\lambda_j\nonumber\\
&&+\mathcal{S}_q(A|B), 
%&=&\mathcal{S}(B)+\mathcal{S}(A,B)+\mathcal{S}_q(A|B),\nonumber\\
\end{eqnarray}
where $\lambda_j$ are the eigenvalues of (\ref{doX}),
\begin{eqnarray*}
\lambda_1&=&\frac{1}{2}\left[  \rho_{11}+\rho_{44} +\sqrt{(\rho_{11}-\rho_{44})^2+
4\left|\rho_{14}\right|^2}   \right],\\
\lambda_2&=&\frac{1}{2}\left[  \rho_{11}+\rho_{44} -\sqrt{(\rho_{11}-\rho_{44})^2+
4\left|\rho_{14}\right|^2}   \right],\\
\lambda_3&=&\rho_{22}+\left|\rho_{23}\right|,\\
\lambda_4&=&\rho_{22}-\left|\rho_{23}\right|.
\end{eqnarray*}

When $\rho_{14}\rho_{23}\neq0$ Eq. (\ref{conde}) provides 
the correct value for the quantum conditional entropy. However, 
for $\rho_{14}\rho_{23}=0$ the correct value of the conditional entropy is not 
always given by this expression \cite{numDIS,valAE}. In the latter case the 
minimization depends on only one single parameter and should be numerically 
carried out. For all the cases studied here, however, the numerical 
calculations agreed with the analytical expression of quantum discord. 

Finally, for the density operator (\ref{doX}) the EoF is given by \cite{Woo98}
\begin{eqnarray}\label{eof}
EoF(\rho_{AB})&=&-g\log_2g-(1-g)\log_2(1-g),
\end{eqnarray}
where $g=(1+\sqrt{1-C^2})/2$ and $C=2\max\left\{0,\Lambda_1,\Lambda_2 \right\}$
is the concurrence with $\Lambda_1=|\rho_{14}|-\sqrt{\rho_{22}\rho_{33}}$ and 
$\Lambda_2=|\rho_{23}|-\sqrt{\rho_{11}\rho_{44}}$.


\begin{thebibliography}{99}

\bibitem{Sac99} S. Sachdev, Quantum Phase Transitions (Cambridge University 
Press, Cambridge, 1999).

\bibitem{rowley} S. Rowley, R. Smith, M. Dean, L. Spalek, M. Sutherland, M. 
Saxena, P. Alireza, C. Ko, C. Liu, E. Pugh, S. Sebastian, and G. Lonzarich, 
Phys. Status Solidi B \textbf{247}, 469 (2010).

\bibitem{dolgopolov} V. F. Gantmakher and V. T. Dolgopolov, Physics-Uspekhi 
\textbf{53}, 1 (2010).

\bibitem{greiner} M. Greiner, O. Mandel, T. Esslinger, T.W. H\"ansch, and I. 
Bloch, Nature (London) \textbf{415}, 39 (2002).

\bibitem{lidar} L.-A. Wu, M. S. Sarandy, and D. A. Lidar, Phys. Rev. Lett. 
\textbf{93}, 250404 (2004).

\bibitem{oliveira} T. R. de Oliveira, G. Rigolin, M. C. de Oliveira, and E. 
Miranda, Phys. Rev. Lett. \textbf{97}, 170401 (2006); 
T. R. de Oliveira, G. Rigolin, and M. C. de Oliveira, Phys. Rev. A \textbf{73},
010305(R) (2006); G. Rigolin, T. R. de Oliveira, and M. C. de Oliveira, Phys. 
Rev. A \textbf{74}, 022314 (2006); T. R. de Oliveira, G. Rigolin, M. C. de 
Oliveira, and E. Miranda, 
Phys. Rev. A \textbf{77}, 032325 (2008).

\bibitem{ortiz} H. Barnum, E. Knill, G. Ortiz, R. Somma, and L. Viola,
Phys. Rev. Lett. \textbf{92}, 107902 (2004); 
R. Somma, G. Ortiz, H. Barnum, E. Knill, and L. Viola, Phys. Rev. A 
\textbf{70}, 042311 (2004).

\bibitem{zurek} H. Ollivier and W. H. Zurek, Phys. Rev. Lett. \textbf{88}, 
017901 (2001); 

\bibitem{Henderson} L. Henderson and V. Vedral, J. Phys. A \textbf{34}, 6899 
(2001).

\bibitem{Dil08} R. Dillenschneider, Phys. Rev. B \textbf{78}, 224413 (2008).

\bibitem{Sar09} M. S. Sarandy, Phys. Rev. A \textbf{80}, 022108 (2009).

\bibitem{werPRL} T. Werlang, C. Trippe, G.A.P. Ribeiro, and G. Rigolin, Phys. 
Rev. Lett. \textbf{105}, 095702 (2010).

\bibitem{nielsen} M. A. Nielsen and I. L. Chuang, Quantum Computation and 
Quantum Information (Cambridge University Press, Cambridge, 2000).

\bibitem{bayes} T. M. Cover and J. A. Thomas, Elements of Information Theory  
(Wiley-Interscience, New York, 2006).

\bibitem{peres} A. Peres, Quantum Theory: Concepts and Methods (Kluwer 
Academic Publishers, New York, 2002).

\bibitem{ved01} B. Dakic, V. Vedral, and C. Brukner, Phys. Rev. Lett. 
\textbf{105}, 190502 (2010).

\bibitem{footnote1} Similarly, we have $D(B|A)=0$ if and only if there exists 
a von Neumann measurement $\left\{\Pi_j^A\right\}$ such that $\rho_{AB}=\sum_j 
\left(\Pi_j^A \otimes \boldsymbol{1}_B\right)\rho_{AB} \left(\Pi_j^A\otimes
\boldsymbol{1}_B\right)$.

\bibitem{cstates} J. Oppenheim, M. Horodecki, P. Horodecki, and 
R. Horodecki, Phys. Rev. Lett. \textbf{89}, 180402 (2002); K. Modi, 
T. Paterek, W. Son, V. Vedral, and M. Williamson, Phys. Rev. 
Lett. \textbf{104}, 080501 (2010).

\bibitem{arnesen} M. C. Arnesen, S. Bose, and V. Vedral, Phys. Rev. Lett. 
\textbf{87}, 017901 (2001).

\bibitem{Wer10} T. Werlang and G. Rigolin, Phys. Rev. A \textbf{81}, 044101 
(2010).

\bibitem{minDIS} S. Hamieh, R. Kobes, and H. Zaraket, Phys. Rev. A 
\textbf{70}, 052325 (2004).

\bibitem{numDIS} D. Girolami and G. Adesso, eprint arXiv:1103.3189v1 [quant-ph].

\bibitem{Wer89} R. F. Werner, Phys. Rev. A \textbf{40}, 4277 (1989).

\bibitem{entangle} A. Peres, Phys. Rev. Lett. \textbf{77}, 1413 (1996).

\bibitem{Woo98} W. K. Wootters, Phys. Rev. Lett. \textbf{80}, 2245 (1998); 
T. Yu and J. H. Eberly, Quantum Inf. Comp. \textbf{7},459 (2007).

\bibitem{discordX} M. Ali, A. R. P. Rau, and G. Alber, Phys. Rev. A 
\textbf{81}, 042105 (2010); ibid. \textbf{82}, 069902(E) (2010).

\bibitem{YANG} C.N. Yang and C.P. Yang,  Phys. Rev. \textbf{147}, 303 (1966).


\bibitem{KLUMPER92} A. Kl\"umper, Ann. Phys. \textbf{1}, 540 (1992); 
A. Kl\"umper, Z. Phys. B \textbf{91}, 507 (1993).

\bibitem{NLIE} M. Bortz and F. G\"ohmann, 
Eur. Phys. J. B \textbf{46}, 399 (2005); 
H.E. Boos, J. Damerau, F. G\"ohmann, A. Kl\"umper, J. Suzuki, and A. Wei\ss e, 
J. Stat. Mech. (2008) P08010; C. Trippe, F. G\"ohmann, and A. Kl\"umper, 
Eur. Phys. J. B \textbf{73}, 253 (2010).

\bibitem{TAKAHASHI} M. Takahashi, Thermodynamics of one-dimensional solvable 
models (Cambridge University Press, Cambridge, England, 1999).

\bibitem{GAUDIN} J. Cloizeaux and M. Gaudin, J. Math. Phys. \textbf{7}, 
1384 (1966).

\bibitem{rulli}  C. C. Rulli and M. S. Sarandy, Phys. Rev. A \textbf{81}, 
032334 (2010).

\bibitem{sudden} J. Maziero, L. C. C\'eleri, R. M. Serra, and V. Vedral, Phys. 
Rev. A \textbf{80}, 044102 (2009).

\bibitem{Rig04} G. Rigolin, Int. J. Quantum Inform. \textbf{2}, 393 (2004).

\bibitem{LSM} E. Lieb, T. Schultz, and D. Mattis, Ann. Phys. \textbf{16}, 
407 (1961); E. Barouch, B.M. McCoy, and M. Dresden, Phys. Rev. A 
\textbf{2}, 1075 (1970);  
E. Barouch and B.M. McCoy, Phys. Rev. A \textbf{3}, 786 (1971).

\bibitem{isingQPT} P. Pfeuty, Ann. Phys. (New York) \textbf{57}, 79 (1970).

\bibitem{anisQPT} M. Zhong and P. Tong, J. Phys. A: Math. Theor. 
\textbf{43}, 505302 (2010). 

%\bibitem{diagXY} E. Barouch, B. M. McCoy, and M. Dresden, Phys. Rev. A 
%\textbf{2}, 1075 (1970); E. Barouch and B. M. McCoy, 
%Phys. Rev. A \textbf{3}, 786 (1971).

\bibitem{osborne} T. J. Osborne and M. A. Nielsen, Phys. Rev. A 
\textbf{66}, 032110 (2002).

\bibitem{Sar10} J. Maziero, H. C. Guzman, L. C. C\'eleri, M. S. Sarandy, and 
R. M. Serra, Phys. Rev. A \textbf{82}, 012106 (2010).

\bibitem{Ami10} B. Tomasello, D. Rossini, A. Hamma, and L. Amico, eprint 
arXiv:1012.4270v1 [quant-ph].

\bibitem{valAE} Q. Chen, C. Zhang, S. Yu, X.X. Yi, and C. H. Oh, 
eprint arXiv:1102.0181v1 [quant-ph].


\end{thebibliography}
\end{document}